\documentclass[]{iopart}
\usepackage{cite,enumerate}
\usepackage{graphicx}
\usepackage{amsfonts,amssymb,bm}
\usepackage[usenames]{color}

\usepackage{fancyhdr}
\begin{document}

\topical[Precursor films in wetting phenomena]
{Precursor films in wetting phenomena}

\author{M N Popescu$^{1,2}$, G Oshanin$^{3}$, S Dietrich$^{2,4}$, and
A-M Cazabat$^{5}$}

\address{$^1$ Ian Wark Research Institute, University of South Australia,
Adelaide, SA 5095, Australia
}
\address{$^2$ Max Planck Institute for Intelligent Systems, Heisenbergstr. 3,
70569 Stuttgart, Germany
}
\address{$^3$ Laboratoire de Physique Th{\'e}orique de la Mati{\`e}re
Condens{\'e}e (UMR CNRS 7600), Universit\'e Pierre et Marie Curie,
4 place Jussieu F-75252, Paris Cedex 05, France
}
\address{$^4$ Institut f\"ur Theoretische und Angewandte Physik,
Universit\"at Stuttgart, Pfaffenwaldring 57, 70569 Stuttgart,
Germany
}
\address{$^5${Laboratoire de Physique Statistique (UMR CNRS 8550), Universit\'e Pierre et Marie Curie and Ecole Normale Sup\'erieure,
24 rue Lhomond 75231 Paris Cedex 05,  France}
}

\eads{\mailto{Mihail.Popescu@unisa.edu.au}, \mailto{oshanin@lptmc.jussieu.fr},
\mailto{dietrich@is.mpg.de}, \mailto{anne-marie.cazabat@upmc.fr}}

\begin{abstract}
The spontaneous spreading of non-volatile liquid droplets on solid substrates
poses a classic problem in the context of wetting phenomena. It is well known
that the spreading of a macroscopic droplet is in many cases accompanied by a
thin film of macroscopic lateral extent, the so-called precursor film, which
emanates from the three-phase contact line region and spreads ahead of the
latter with a much higher speed. Such films have been usually associated with
liquid-on-solid systems, but in the last decade similar films have been
reported to occur in solid-on-solid systems. While the situations in which the
thickness of such films is of mesoscopic size are rather well understood, an
intriguing and yet to be fully understood aspect is the spreading of
microscopic, i.e., molecularly thin films. Here we review the available
experimental observations of such films in various liquid-on-solid and
solid-on-solid systems, as well as the corresponding theoretical models
and studies aimed at understanding their formation and spreading dynamics.
Recent developments and perspectives for future research are discussed.
\end{abstract}


\tableofcontents

\section{\label{sec_intro}Introduction}

Wetting of a solid by a liquid is a very common natural phenomenon.
Morning dew on the grass, rain drops on the plant leafs or windows,
the coffee spill staining the table cloth, the water rise in the
capillaries of a tree, the lubricating film covering the eye, or the oil
drops on a non-sticky frying pan are examples of wetting in day-to-day life.
As much as being a ubiquitous natural phenomenon, wetting is at the core of
many technologies and technological processes vital for various
industries \cite{Padday1978,Quere,Starov}, e.g., the protective spin coating
of surfaces (CDs, DVDs, glass lenses, car mirrors and windows), the
development of water-resistant fabric, ink-jet printing, wall-painting,
froth flotation  \cite{Fuerstenau2007} or acid heap
leaching \cite{Michaelis1989} in minerals recovery.

The theoretical analysis of wetting phenomena has been started more
than two hundred years ago by Young \cite{Young}, Laplace \cite{Laplace}
and Plateau \cite{Plateau}. Their descriptions and the related subsequent
developments are usually referred to as classical capillarity.
Based on the seminal work by Gibbs \cite{Gibbs1957}, the theory of molecular
liquids in terms of statistical physics led to a good (and still progressing)
understanding of their bulk equilibrium properties \cite{Widom1982,Hansen1986}.
The theoretical description of the structure and properties of free
liquid-vapor interfaces \cite{Evans1979,Evans1990,Dietrich1993,Mecke1999},
as well as of the equilibrium wetting transitions occurring at an inert wall
exposed to a fluid which is in a thermodynamic state close to bulk
liquid-vapor coexistence, have also reached a mature state
\cite{deGennes1985,daGama1986,Dietrich1988,Schick1990}. (For rather recent
developments see Refs. \cite{Parry2011,Parry2009,Parry2004,Binder2011} and
references therein.)
\footnote[1]{
\label{foot1}
Unfortunately size constraints make it impossible to do justice to all the
numerous important contributions to the understanding of equilibrium wetting
phenomena. The interested reader is encouraged to consult the extended
bibliography lists in, e.g., Refs. \cite{Evans1990,Dietrich1988,deGennes1985}.}

These contributions have put forward a thermodynamic and mechanical
description of capillarity which is successful in explaining a large number
of phenomena such as capillary rise, the shape of sessile or
pendant drops and the shape of a meniscus, at least as long as only
macroscopic phenomena are involved. The physico-chemical parameters
controlling the thermodynamic wettability of solid surfaces were
clarified through the careful work of Zisman \cite{Zisman}
and colleagues (see, e.g., Ref. \cite{Padday1978}).

In the last two decades, significant efforts have been made towards
understanding wetting phenomena at mesoscopic and microscopic scales.
This was facilitated by the development of modern experimental techniques,
capable of probing interfacial structures down to molecular scales, together
with the fast paced technological progress in the fabrication of patterned
surfaces possessing lateral wetting properties tailored on the micron- or
even nano-scale through surface chemistry or topography. In this review, we
focus solely on one particular subject in this area, i.e., the spreading of
molecularly thin precursor films emanating from droplets deposited on
solid substrates.

Since the late 1980s much progress has been achieved, both experimentally
and
theoretically, towards the understanding of the spreading dynamics of such
films. Not only their occurrence in various liquid-on-solid systems has
been reported in numerous studies, but also various details about their
spatio-temporal structure have been revealed. Moreover, molecularly thin
precursor films have been discovered in metal-on-metal systems, including
even the case of films emerging out of solid clusters. The rich behaviour
observed in these latter systems, which includes surface alloying, is at
present a topic of active investigations. The rapidly developing nano-fluidics
chip technology keeps the interest in such films high by raising new questions,
such as the spreading behaviour of sub-micron scale drops or their motion
induced by such precursor films \cite{Zhao2011}. On the theoretical side, a
number of models have been proposed in order to explain the experimentally
observed behaviour.

However, these results are scattered in the literature and up to date no
attempt has been made to present a comprehensive review. In particular
Ref. \cite{Joanny1992}, which dates back to 1991, covered only the few
results concerning the dynamics of microscopic precursor films available
during the short period since 1989, when the first reports on this topic
appeared \cite{cazabat1,beaglehole,cazc,cazabat2}.
The recent review of dynamics of wetting in Ref. \cite{Bonn2009} has only
briefly touched this subject because it was not within its main scope.
Here we strive for overcoming this dearth by collecting and assessing the
large body of available experimental results, both for liquid-on-solid
and solid-on-solid systems, and by discussing the various attempts of
theoretical and numerical modeling the spreading of microscopically thin
precursor films.

This review is outlined as follows. In Section \ref{par_tot} we briefly
recall certain general thermodynamic criteria for the thermodynamic
equilibrium of a macroscopic (but sufficiently small for the capillary
forces dominating external ones, such as gravity) liquid drop on a solid
substrate. In Section \ref{spon} we focus on spontaneous spreading of a
droplet. In order to put the subject into a broader context (and in order to
avoid referring an interested reader to the seminal original papers too
often), there we concisely discuss the characteristic features of the spreading
of the macroscopic part of the droplet, the emergence of a mesoscopic precursor
film, and its dynamics. Size constraints do not permit to provide a more
detailed discussion of all seminal contributions. An interested reader can
find such a description in comprehensive reviews
\cite{deGennes1985,Joanny1992,Bonn2009}.
Next, in Section \ref{sec_exper} we turn to the subject of interest here and
we overview  experimental data on spreading of microscopically thin precursor
films in different liquid-on-solid and solid-on-solid systems. Section
\ref{sec_models} presents various theoretical models of microscopic
precursor films, while Section \ref{sec_num} focuses on numerical
simulations of these systems. Section \ref{sec_new} presents the recent
developments in this area. Finally, in Section \ref{summary} we provide
a summary and an outlook.

\section{\label{par_tot} Droplets on solid substrates: equilibrium versus
non-equilibrium}

One of the often encountered realizations of wetting is that of a liquid
drop on a planar solid substrate. For this setup, two limiting cases are
well-defined:
(i) a thermodynamic equilibrium situation
\footnote[2]{
\label{foot2}
There is mechanical equilibrium, chemical equilibrium (i.e., the matching of the
chemical potential of each species present in the fluid so that the phases are
mutually saturated) and thermal equilibrium (i.e., temperature matching) between
liquid and gas, so that the gas is the saturated vapour of the liquid. The
simultaneous occurrence of mechanical, chemical and thermal equilibrium is referred
to as thermodynamic equilibrium.
}
with a liquid droplet on a solid substrate and the latter covered by a
thin film, both in contact with the saturated coexisting vapour phase in a bounded system,
and (ii) a non-equilibrium situation with late-stage spreading (so that all
transient effects, e.g., the ones due to the initial shape, have died out) of a
droplet of a non-volatile liquid on an otherwise ``dry'' solid substrate.

\subsection{Partial wetting}

\subsubsection{Thermodynamic equilibrium.}
\begin{figure}[!htb]
\includegraphics[width=.46 \linewidth]{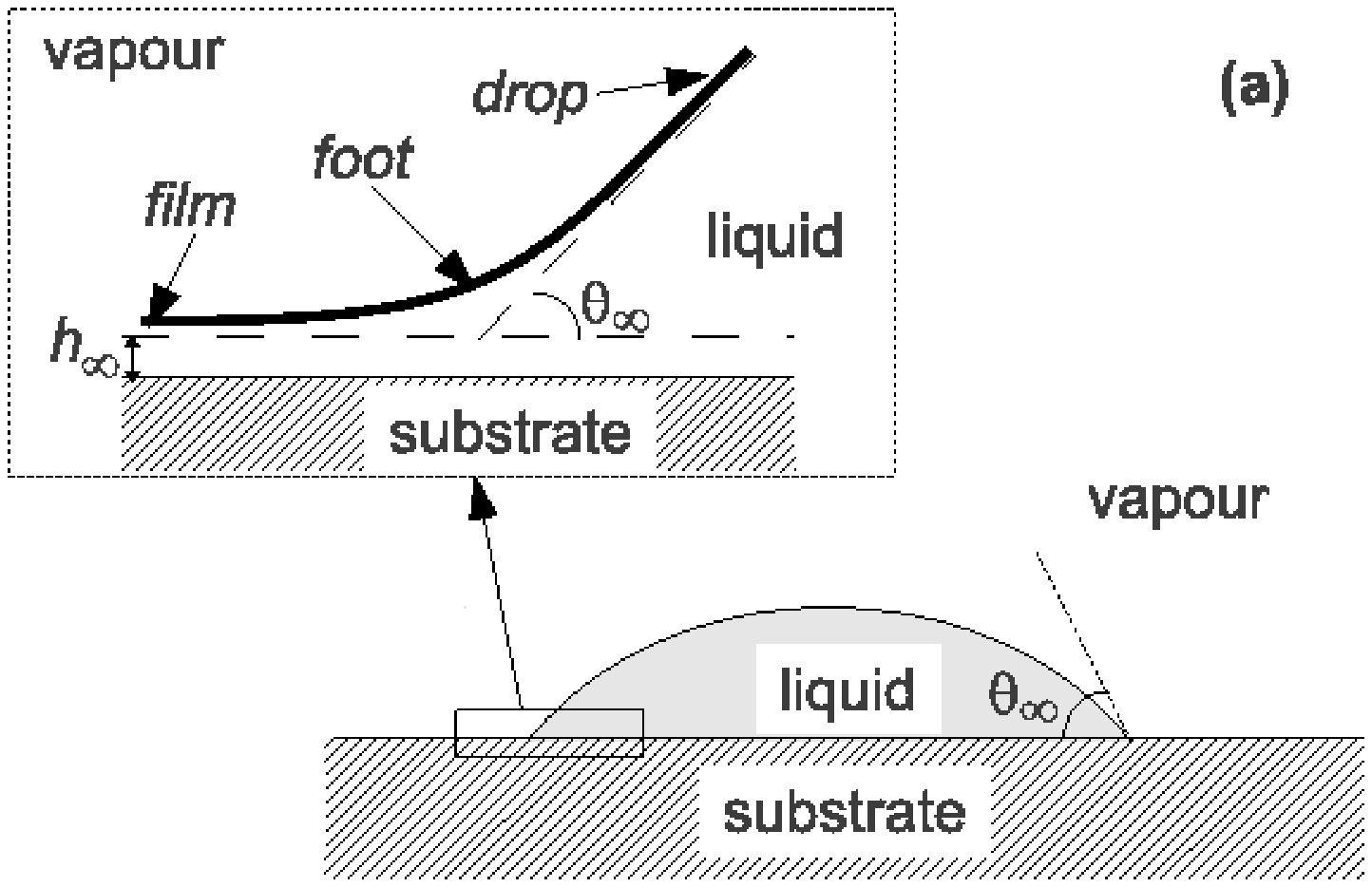}%
\hspace*{.03 \linewidth}%
\includegraphics[width=.51 \linewidth]{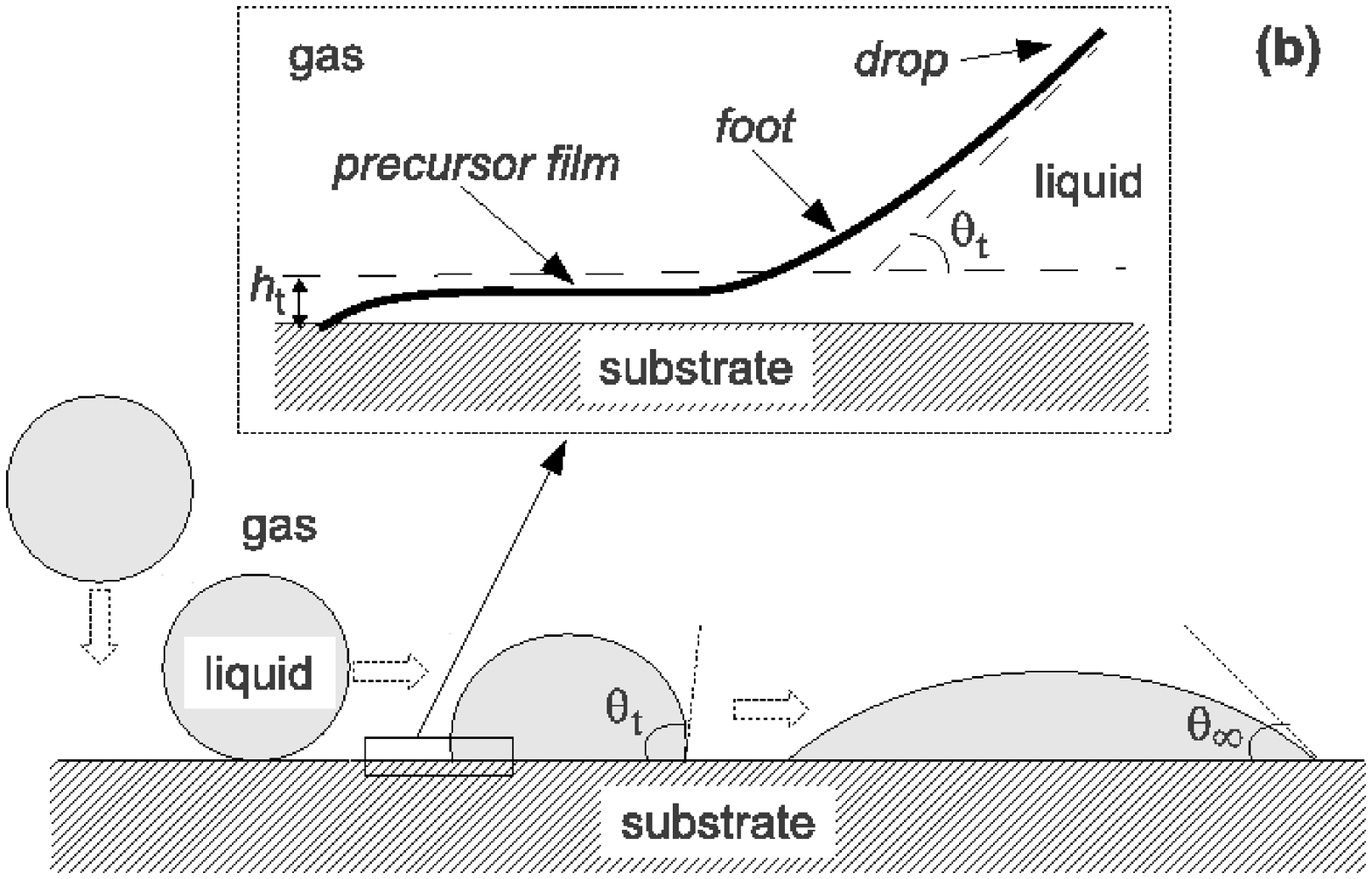}
\caption{
\label{fig1}
{\bf (a)} Schematic drawing of a liquid droplet in thermal equilibrium with its
saturated vapour and in contact with an inert, flat substrate covered by a
thin film of thickness $h_\infty$ at a temperature $T$ below the wetting
transition $T_W$ in a bounded system. The inset zooms into the zone near
the three-phase contact line and emphasizes that the macroscopic spherical
cap is connected with a film covering the substrate via a ``foot''
region of mesoscopic size (with a height less than 100 nm) where the shape of
the liquid-vapour interface is determined by the effective interface potential.
{\bf (b)} Schematic drawing of a spherically shaped droplet of a non-volatile
liquid spreading on an inert, flat and unbounded substrate. The inset shows
the precursor film emerging during the spreading.
}
\end{figure}

In the thermodynamic equilibrium situation, three distinct interfaces meet at
a contact line [see Fig. \ref{fig1}(a)]: the liquid-vapour, the liquid-solid
and the solid-vapour interfaces with the corresponding well-defined surface
tensions  $\sigma_{lv}$, $\sigma_{ls}$, and $\sigma_{sv}$, respectively.
Note that $\sigma_{sv}$ takes into account the presence of a thin liquid-like
wetting film intervening between the solid substrate and the vapour phase
\cite{Dietrich1988}.
In the case of a macroscopically large drop (but with a linear size smaller
than the capillary length $\chi = \sqrt{\sigma_{lv}/(\rho g)}$, where
$\rho$ is the mass density of the liquid and $g$ the gravitational constant) on
a perfectly smooth, chemically homogeneous, and planar substrate (to
which we shall constrain most of our discussion) the equilibrium contact
angle
$\theta_{\infty}$
[see Fig. \ref{fig1}(a)] is determined by Young's equation
\begin{equation}
\label{Young}
\sigma_{lv} \cos \theta_{\infty} = \sigma_{sv}-\sigma_{sl}.
\end{equation}
Note that $\theta_\infty$ is understood to be measured macroscopically, on
a scale large compared with the range of long-ranged intermolecular forces.
In terms of the so-called spreading coefficient (also often called the
spreading ``power'')
\begin{equation}
S = \sigma_{sv} - \sigma_{lv} - \sigma_{sl},
\end{equation}
one can rewrite Eq.~(\ref{Young}) as
\begin{equation}
\cos \theta_{\infty}  = 1 + \frac{S}{\sigma_{lv}},
\end{equation}
which relates the equilibrium contact angle to $S$.

In thermal equilibrium $S$ is either negative or zero
(Antonov's rule \cite{Dietrich1988}). The case $S < 0$, which implies
$0^\circ < \theta_{\infty} \leq 180^\circ$, is referred to as partial wetting.
It corresponds to drops, surrounded by a microscopically thin film adsorbed at
the surface. Classical capillarity predicts the essential features of the shape
of such drops. If the drop is small enough so that gravity can be neglected
(i.e., the characteristic length scale of the drop is much smaller
than the capillary length $\chi$), the hydrostatic pressure rapidly
equilibrates inside the drop; the drop then adopts the shape of a
spherical cap in order to obey Laplace's law (i.e., the pressure difference
between the liquid drop and the surrounding vapour is balanced by the
surface tension $\sigma_{lv}$ multiplied by the mean curvature of the drop
shape) and the in-plane isotropy.
\footnote[3]{
\label{foot3}
Note, however, that the contact angles of partially wetting
droplets are very sensitive to various surface defects, surface roughness,
contaminants or chemically active sites present on the substrate
\cite{deGennes1985,Joanny1992,Bonn2009}.
}

Statistical mechanics studies of wetting phenomena in the grand canonical
ensemble reveal that in the case of partial wetting (i.e., at bulk
liquid-vapour coexistence on the vapour side and at a temperature $T$ below
the wetting transition temperature $T_w$) the substrate is covered by a
microscopically thin liquid-like film the thickness
$h_{\infty}$ of which [see the inset in Fig. \ref{fig1}(a)] is temperature
dependent and exhibits wetting
transitions \cite{Dietrich1988,Pandit1983}. In the case of a first-order
wetting transition, for $T \to T_w$ $h_{\infty}$ jumps to a macroscopic value,
whereas $h_{\infty}$ diverges continuously for a second-order wetting
transition (also known as critical wetting \cite{Dietrich1988}). For bounded
systems these singularities are smeared out. The spherical cap-like shape of
the liquid-vapour interface, which for a suitably large drop is indeed valid
far away from the substrate, is distorted in the vicinity of the substrate
due to the effective interaction between the liquid-vapour interface and the
substrate and smoothly connects to the uniform equilibrium film thickness far
away from the drop \cite{Dietrich1988,Bauer1999}. The deviations of this
actual shape from that of a spherical cap cut by the substrate contribute to
the line tension
\cite{Bauer1999,Indekeu1994,Getta1998,Dobbs1999,Pompe2002,Herminghaus2000,
White1977,Churaev1982,Schimmele2007}, i.e., the excess free energy,
associated with a line-like inhomogeneity, within the
decomposition of the excess
free energy into bulk, surface and line contributions \cite{Gibbs1957}.
\footnote[4]{
\label{foot4}
Instabilities of droplets caused by line tension have been
critically discussed in Ref. \cite{Mechkov3}.
}

\subsubsection{Non-equilibrium behaviour.}
If a droplet is deposited on an initially dry substrate, it is typically not
in equilibrium and the equilibrium can be reached only via certain dynamical
processes. The way the relaxation towards equilibrium takes place depends
significantly on the volatility of the liquid under ambient conditions. If
the liquid is volatile, the equilibrium with the ambient gas can be
established relatively fast via condensation of vapour onto the substrate. On
the contrary, if the liquid is non-volatile under ambient conditions (i.e.,
if it has a very low vapour pressure), the evolution of the drop shape
towards equilibrium will proceed via two-dimensional evaporation and motion
of liquid molecules on the substrate surface. Usually, these are much slower
processes and in many cases the time required to approach equilibrium after
contact with the ``dry'' substrate can be more than a few hours, or even days.

The pertinent quantity for the dynamics is the initial spreading coefficient
\begin{equation}
S_0 = \sigma_{sg} - \sigma_{lg} - \sigma_{sl},
\end{equation}
where $\sigma_{sg}$
is the surface tension of the dry solid substrate in contact with the gas.
The surface tension $\sigma_{sg}$ is typically larger than $\sigma_{sv}$;
assuming that $\sigma_{lg} \simeq \sigma_{lv}$, which is appropriate
for an inert gas (i.e., which has a low solubility in the liquid and which
does not act as a surfactant) and for temperatures sufficiently far from the
critical point, it follows that $S_0 \geq S$. (Note that $\sigma_{sg}$ is not a
measurable quantity; it can be computed assuming the constrained equilibrium that
the gas density is constant up to the substrate surface without the formation of a
liquid-like wetting film.) The initial $S_0$ can be positive or negative,
but whenever $S_0 < 0$, $S$ is also negative. Hence, for non-volatile liquids
and for $S_0 < 0$ the drop will relax toward an equilibrium configuration
which displays a non-zero static (equilibrium) contact angle.

\subsection{Complete wetting}

If the equilibrium spreading coefficient vanishes, i.e., $S = 0$, one has
complete wetting. The corresponding equilibrium contact angle of a volatile
liquid is zero, i.e., $\theta_{\infty} = 0$, so that the solid-vapor interface
consists of a macroscopically thick wetting layer, which implies that the
solid-vapour surface tension equals the sum of the solid-liquid and
liquid-vapour surface tension.

For a drop of a non-volatile liquid deposited on a initially dry substrate
$S_0$ is positive (although its equilibrium spreading power $S = 0$), and
therefore the drop will spread while preserving its volume. This volume
conservation constraint is essential for describing both the spreading dynamics
and the final state of non-volatile liquid drops. It explains why in certain
situations a non-volatile liquid drop does not spread down to an unbounded
film of uniform thickness, but that spreading stops when the
drop eventually takes on a ``pancake''-like shape of thickness
$e(S_0) = a \sqrt{3 \,\sigma_{lg}/ (2\, S_0)}$ (where $a$ is a molecular
length \cite{deGennes1985} specific to nonretarded van der Waals interactions,
$a = \sqrt{- A/(6\, \pi\, \sigma_{lg})}$, with
$A < 0$ being the Hamaker constant). This occurs if short-ranged
interactions promote dewetting, even though the overall situation is that of
complete wetting. Such structures have been theoretically predicted and
analyzed in Refs. \cite{deGennes1985,Ruckenstein,Joanny2}, and indeed
observed later experimentally (see, e.g., Ref. \cite{cazabat4}).

\section{\label{spon}Spreading of non-volatile droplets}

The issue of a non-volatile drop spreading on a substrate from an initial
non-equilibrium configuration with $S_0 > 0$ towards its equilibrium shape
in situations in which $S = 0$ has been the subject of extensive experimental
and theoretical investigations. The results of this analysis have been
reviewed thoroughly in Refs.
\cite{deGennes1985,Cazabat1987,Cazabat1991,Joanny1992,Bonn2009,Rauscher2008}.
Here we shall focus only on those main findings which are of interest in the
present context. Moreover, we constrain ourselves to the late stages of
spreading (and hence, sufficiently small dynamical contact angles) when all
transient behaviours and memory-effects of the initial non-equilibrium shape
have died out. (Some recent experimental results on the transient inertial
regime at very early stages of spreading can be found in Refs.
\cite{Bird2008,Chen2011}.)

In  Fig. \ref{fig1}(b) we sketch a typical configuration for a non-volatile
liquid droplet spreading on a solid substrate. The drop can be divided into
the following two main regions: (i) a 'macroscopic' bulk and
(ii) a precursor film, the thickness of which may be on a mesoscopic or a
microscopic scale. Note that Fig. \ref{fig1}(b) is schematic and the relative
sizes of these regions are not drawn to scale.
The \textit{non-equilibrium, time-dependent} parameters $\theta_t$ and $h_t$,
which are expected to obey
$\theta_t (t \to \infty) \to \theta_\infty$ and $h_t (t \to \infty) \to
h_\infty$, require further discussion.
During spreading the shape of the macroscopic part of the
drop does not necessarily has to be a spherical cap because spreading is a
non-equilibrium process. (Although very often a spherical cap shape is
employed in the interpretation of experimental or computer simulations
results concerning droplet spreading or dynamic wetting  (see, e.g.,
Ref. \cite{Blake1997}) one should keep in mind that it remains an
assumption which should be at least checked for validity in each
particular instance.) Therefore we shall define the parameters $\theta_t$
and $h_t$ without reference to a particular shape. As shown in
Fig. \ref{fig1}(b), far away from the substrate the shape of the drop has a
negative curvature, while at the precursor film the curvature is
approaching zero from above. Therefore there is an inflection point, at a
distance from the substrate of the order of the range of the effective
interface potential, where the curvature of the shape changes from negative
(corresponding to the macroscopic part) to positive (corresponding to the
region of the foot and precursor film). Therefore, it is a natural choice
to define the macroscopic part of the drop as that part of the shape
exhibiting negative curvature up to the inflection point and the
``dynamic'' contact angle $\theta_t$ as the angle formed by the tangent
to the drop shape at this inflection point (see also Ref. \cite{Mechkov1}).
The foot of the drop is defined as that part of the drop shape where
the curvature is positive and attaining zero from above upon approaching
the almost flat precursor film. Accordingly, $h_t$ is defined as
the maximum thickness of the film extending ahead of the foot region.
Note that for very small or very flat drops (i.e., small equilibrium
contact angles) these definitions involving a dynamic macroscopic shape of
the drop and a foot region are obviously of limited value because they
increasingly depend on the \textit{experimental resolution} and, moreover,
the inflection point becomes inadequate as a relevant experimental parameter.
In contrast, the definitions of the equilibrium observables $\theta_\infty$ and
$h_\infty$ remain on solid grounds as long as the amount of liquid is
sufficient to limit the fluctuations around a mean shape.

\subsection{Non-volatile droplet spreading at macroscopic scales}
\label{macroscopic}

The key parameter characterizing the spreading of the macroscopic part of
the non-volatile drop is the so called capillary number
${\rm Ca} = \eta U/\sigma_{lg}$, where $U$ is the velocity of the advancing
liquid wedge and  $\eta$ is the viscosity of the liquid. An early experimental
analysis of the relation between the dynamical contact angle $\theta_t$ and
the capillary number has been carried out by Hoffman \cite{Hoffman}, who
studied a forced spreading (with velocities varying over five decades) of a
liquid in a capillary, measuring a dynamical contact angle by a photographic
technique. In a series of experiments with silicon oils, he obtained conditions
of complete wetting and found a rather universal relation between ${\rm Ca}$
and $\theta_t$; in particular, in the limit of low ${\rm Ca}$ and $\theta_t$
he found that his data can be represented as a power law
\begin{equation}
\label{ca}
{\rm Ca} = {\rm constant} \, \theta_t^{\alpha},
\end{equation}
where $\alpha = 3 \pm 0.5$. Cox \cite{Cox} treated analytically the rather general
problem of one fluid displacing another one on a solid surface for any viscosity
ratio and contact angle. He solved the hydrodynamic equations to first order in
${\rm Ca}$ assuming that the macroscopic wedge with angle $\theta$ ends in a zone of
molecular extension where slip is allowed. In the complete wetting case and for
small contact values his analysis confirmed the result in Eq. (\ref{ca}) with
$\alpha = 3$. We note that a relation of this type was first derived
theoretically by Fritz \cite{Fritz}, but for the slightly different problem of a
liquid spreading on a wet surface. In comparison with Cox's analysis, in this case
the viscosity of the displaced phase (i.e., vapour) is negligible and the conceptual
difficulties
arising due to the movement of the contact line on a solid substrate
are bypassed by considering a surface covered by a liquid film. A detailed
analysis of the relation in Eq.(\ref{ca}) has been carried-out also by
Teletzke \textit{et al} \cite{Teletzke}.

Voinov \cite{Voinov} and Tanner \cite{Tanner} derived analytically the relation
in Eq. (\ref{ca}) with $\alpha = 3$ solving the hydrodynamic equations in the
lubrication approximation. They have also used the result in Eq. (\ref{ca}) in order to
obtain the time evolution of the base radius $R_t$ and the dynamical contact
angle $\theta_t$ of a spreading droplet of a non-volatile liquid
under the assumption that the droplet shape is that of a spherical cap at all
times during spreading. In this case $R_t$ and $\theta_t$ are coupled due to
the constraint of fixed volume $V$.
This leads to the so-called Tanner's laws \cite{Voinov,Tanner} according to which
for sufficiently large $R_t$ and small $\theta_t$ these quantities are given by
\begin{equation}
\label{tanner}
R_t \sim V^{3/10}
\left(\frac{\sigma_{lg}}{\eta} \,t\right)^{1/10}
\end{equation}
and
\begin{equation}
\label{con}
\theta_t \sim V^{1/10}
\left(\frac{\sigma_{lg}}{\eta} \,t\right)^{-3/10}.
\end{equation}
Thus the spreading of the macroscopic part of the droplet is rather slow; it
would take a very long time to completely spread a macroscopic drop.
Note that Eqs. (\ref{tanner}) and (\ref{con}), together with the fixed
volume constraint, predict that the final state (i.e., at $t \to \infty$) of the
spreading drop is a two-dimensional gas. However, this is not necessarily so. The
validity of the relations (\ref{tanner}) and (\ref{con}), which are based on classical continuum hydrodynamics,  breaks down at a
crossover time $T_0$ at which the height of the spreading drop reaches the range of
the effective interface potential (which describes the effective interaction
between the substrate and the liquid-gas interface).

The behaviours given in Eqs. (\ref{tanner}) and (\ref{con}) have been
observed in a number of thorough experimental studies (see, e.g., the detailed analysis
of the full shape of a spreading drop by using laser light interferometry reported
by Chen and Wada \cite{Chen} and the review by Marmur \cite{Marmur}). On the
other hand, some significant deviations have been observed for spreading liquids
with very low viscosity
(see, e.g., Ref. \cite{Sawicki}). However, such deviations have been explained in
Ref. \cite{deRuijter} by the influence, at intermediate times, of the dissipation
at the macroscopic contact line \cite{Blake} which is unaccounted
for in the hydrodynamic description of spreading.

We finally point out the remarkable feature in Eqs. (\ref{tanner}) and (\ref{con})
that neither the radius nor the dynamic contact angle of the macroscopic part of the
droplet depend on the initial spreading power $S_0$. This tells that in
conditions in which the liquid completely wets the solid the macroscopic spreading
turns out to be independent of the wettability (i.e., the value of $S_0$) of the
solid surface.
A consistent explanation for this feature is due to Hervet and de
Gennes \cite{Hervet,deGennes1985} who have shown that (i) the macroscopic behaviour described by Eqs. (\ref{ca}), (\ref{tanner}) and (\ref{con}) stems from the interplay of the
hydrodynamic dissipation in the wedge of the macroscopic part of the droplet and
the ``driving force'' for spreading provided by the free energy of the drop due to
its shape not being the equilibrium one, i.e., a two-dimensional gas or a
flat ``pancake'' (see, c.f., Sec. \ref{theor_meso}) and that (ii) $S_0$ is entirely
dissipated in the mesoscopic part of the drop. These arguments have been used
further in Refs. \cite{Mechkov1,Mechkov2} in order to account for the effects of
the line tension on the late-stage (i.e., post-Tanner) spreading of a macroscopic
droplet. For simple liquids, the conclusion reached in these references was that in the late stages of
spreading a positive line tension is responsible for the formation of pancake-like
structures, whereas a negative line tension tends to lengthen the contact line and
to induce an accelerated spreading (i.e., there is a crossover to a faster algebraic increase of $R_t$
than in the Tanner stage) \cite{Mechkov1}.

\subsection{Non-volatile droplet spreading at mesoscopic scales: mesoscopic films}

\subsubsection{\label{exper_meso}Experimental evidence.}

The first reported observation of an ``invisible'' film spreading ahead of the
edge of a macroscopic drop stems from the pioneering work by Hardy
\cite{hardy,hardy1}, who studied the behaviour of drops of water, acetic acid,
and various other polar organic liquids on clean surfaces of glass and steel.
Hardy realized that a film of liquid about one micron thick is pushed out and
spreads from a drop, and that, importantly, this process may or may not be
followed by spreading of the drop itself.
\footnote[5]{
\label{foot5}
More than half a century later, the observation that the spreading
of a drop of water or ethanol depends on the distance from the edge of the
substrate led Marmur to the same reasoning that this can only be caused by a
very thin film spreading ahead of the drop and interacting with the edge
\cite{Marmur1980}.
}

With the experimental techniques available at that time, a direct observation
of such a film -- called by Hardy the "primary" film -- was not possible. He
was able to detect its presence only indirectly, via its lubricating effects,
by observing a significant drop of the static friction of the surface both far
away and close to the three-phase contact line. Hardy stated that he is unable to
conceive a mechanism by which the film is pushed out of the drop and proposed that
spreading  of the film occurs via a process involving a steady condensation of
vapour.
\footnote[6]{
\label{foot6}
Almost 70 years later, clear evidence for an evaporation-condensation
mechanism was reported in Ref. \cite{Novotny1991}, where a microscopic
film was detected on a plate physically separated by a narrow gap from the
substrate with the sessile drop in a partial wetting state. The thickness
profile of this film as a function of the distance from the edge of the
droplet was shown to be identical to that of an uninterrupted film formed
from a droplet residing on this separated plate. This shows that in this case
two-dimensional evaporation out of the three-phase contact line is negligible
compared with three-dimensional evaporation.
}

Two decades later Bangham and Saweris \cite{bang}, analysing the spreading
behaviour of numerous polar and apolar liquids on freshly cleaved mica, have
demonstrated that primary films do also occur even in the absence of any vapour in
the gas. This suggests that the evaporation/condensation scheme
(proposed by Hardy
and clearly confirmed by Ref.\cite{Novotny1991}) is not the only possible physical
mechanism producing the film. The conclusion drawn from their experiments is that
the primary film can also form by surface diffusion of molecules from the edge of
the drop. Recently these findings have been re-examined and confirmed
\cite{Bahadur2009}. The experiments in Ref. \cite{Bahadur2009} employed pairs of
drops of water/alcohol, water/acetic acid, and PDMS/trans-decaline, the first
two involving volatile liquids, placed on glass or mica substrates. The occurrence
of drop motion has been explained as the result of a surface tension gradient at
the solid-gas interface due to the formation of
precursor films \cite{Bahadur2009,Tadmor2009}, which, in case of volatile components, 
form either via an evaporation-condensation mechanism or by
spreading on the surface out of one of the drops (PDMS). It was also noted that the
thickness of these films is different (although no precise characterization of the
thickness was provided): ultra-thin (invisible) in the first case, and mesoscopically
thick (visible) in the latter.

In 1964, Bascom \textit{et al} \cite{bas} investigated the spreading of the primary
film from a more quantitative point of view using ellipsometry and
interferometry techniques. They examined the behaviour of various hydrocarbons
on clean, horizontal or vertical metal surfaces in the presence of either vapour
saturated and vapour unsaturated air. By
monitoring the late stages of spreading via ellipsometry it was concluded that a
primary film is present in all the cases studied. Making the air saturated or
unsaturated with vapour, roughening the surface, and purifying the liquids did
not eliminate the occurrence of the film, but only affected the spreading
speed.
The thickness of the film was found to depend sensitively on the kind of
liquid/solid pair under study but generally it amounted to a few hundred \AA.

A decade later Radigan \textit{et al} \cite{radigan} have shown that the appearance
of such films is not specific to the drop being \textit{bona fide} liquid.
Scanning electron microscopy studies of an inorganic molten glass spreading on
a Fernico metal (i.e., a Fe-Ni-Co alloy with a coefficient of linear expansion
close to
that of hard glass) at $1000^\circ$ C revealed a spreading film of an average height
of the order of one micron. To our knowledge, this was apparently the first
publication in which the notion of a ``precursor film'' has been used.

Finally, Ghiradella \textit{et al} \cite{ghiradella} detected a mesoscopic precursor
film for a hydrochloric solution rising in a glass vessel on a vertical wall.
The existence of the film was demonstrated by monitoring the changes
in electrical resistance across the wall. Ausserr{\'e} \textit{et al} \cite{ausserre}
were the first to directly visualise the precursor film with thicknesses
of several hundred \AA~by using polarised reflection microscopy for the spreading
of non-volatile high-molecular-weight polydimethylsiloxane (PDMS) on smooth
horizontal silicon wafers. The density profiles of these films have been studied
by ellipsometry in Ref. \cite{leger} and the final stages of spreading of small
PDMS droplets in Ref. \cite{daillant}.

\subsubsection{\label{theor_meso}Theoretical concepts.}

We now turn to the theoretical analysis of the time evolution of the
mesoscopic part of the drop (see Fig. \ref{fig1}(b)), focusing on scales from
ca. 30~\AA~to 1
$\mathrm{\mu m}$. At such scales, a continuum picture is still applicable,
but certain long-ranged forces become relevant, mainly van der Waals forces
for organic liquids, or double-layer forces for water. Accordingly, interfacial
tensions alone become insufficient to describe the free energy of the system.
An additional free energy term $\omega(h)$ has to be included,
which takes into account the interactions between the two interfaces
(solid-liquid and liquid-gas for the liquid-on-solid spreading); $h$ is the local film
thickness. This additional free energy contribution, known as the effective
interface potential $\omega(h)$ \cite{Dietrich1988} has a pressure counterpart,
which is the disjoining pressure $\Pi(h) = -\partial \omega /\partial h$. This
was introduced by Derjaguin
\cite{Derjaguin1955} in order to describe the dynamics of thin liquid films.
\footnote[7]{
\label{foot7}
Note that $S_0 = \int^{\infty}_0 dh \, \Pi(h)$, while
$S = \int^{\infty}_{h_{\infty}} dh \, \Pi(h)$, where $h_\infty $ is the
equilibrium film thickness.
}

Within the lubrication approximation \cite{Derjaguin1955,Churaev1982,Oron1997},
the spreading of a thin, non-evaporating liquid film on a solid substrate
ahead of the macroscopic edge of the drop, which moves with
velocity $U$ along the $x$- direction, is governed by the conservation law
\begin{equation}
\label{general1}
\frac{\partial h}{\partial t} = - \frac{\partial }{\partial x} \left(h U\right),
\end{equation}
and the dynamic equation
\begin{equation}
\label{general2}
\eta U = \frac{h^2}{3} \frac{\partial }{\partial x} \left(\sigma_{lg}
\frac{\partial^2 h}{\partial x^2} + \Pi(h) \right),
\end{equation}
which holds for mesoscopically thin films in the absence of gravity and surface
tension gradients.

Capitalizing on the ideas of Derjaguin \cite{Derjaguin1955}, Hervet, de Gennes
and Joanny \cite{Hervet,deGennes1985,Joanny1992,Joanny3,Joanny4} and
Teletzke \textit{et al} \cite{Teletzke} have provided a description of mesoscopic
precursor spreading which includes the contribution of the long-ranged forces.
This has stimulated a strong activity in the field, both theoretically and
experimentally. Following the analysis in Ref. \cite{Joanny4}, below we
shall briefly outline some basic theoretical concepts and results concerning
the spreading of
mesoscopic precursor films.  According to this analysis, explicit results
can be obtained in two limiting cases: first, if the velocity $U$ of the
macroscopic
edge varies very slowly in time (adiabatic case, in which the drift of the
edge
contributes to the spreading of the precursor), and second, if $U \to 0$
(diffusive
case, in which the motion of the macroscopic edge has stopped and the
expansion of
the film is controlled solely by the spreading power $S_0$ and the disjoining
pressure).

\textbf{(A)} {\it Adiabatic films.}
The theoretical analysis of Eqs. (\ref{general1}) and (\ref{general2}) tells that
a so-called adiabatic film  emanates from a liquid wedge which is advancing with a
constant (or slowly varying) velocity $U$. Under the additional
assumptions that
curvature effects are negligible and that this film is stationary, i.e., that
its shape reaches a stationary shape very rapidly at time scales which are
much shorter than the scales at which the velocity of the macroscopic wedge varies
(thus the name ``adiabatic''), the
spreading equation [Eq. (\ref{general2})] of the film along the $x$-coordinate
reduces to \cite{Derjaguin1955,deGennes1984a,Joanny4}:
\begin{equation}
\label{se}
\eta U =  \, \frac{h^2}{3} \frac{\partial \Pi}{\partial x} = {\rm constant}\,,
\end{equation}
which renders the film profile.
The latter equation can be solved, e.g., for the common case of a
non-retarded van der
Waals disjoining pressure $\Pi(h) = -A/(6 \pi h^3)$, where $A < 0$ denotes
the Hamaker constant.
For this case a simple analysis \cite{Joanny4} shows that for $S_0 > 0$
the distance $l_a$ from the macroscopic wedge to the point at which the
height of the adiabatic film is that of a ``pancake'', i.e.,
$h(l_a) = e(S_0) = \sqrt{- A/(6 \pi S_0)}$, equals
\begin{equation}
\label{ad}
l_a = \frac{- A}{6 \pi \eta U e(S_0)} \sim \sqrt{S_0}.
\end{equation}
Hence, the lateral size of the adiabatic film is proportional to $\sqrt{S_0}$. This
shows that, while the macroscopic properties of a spreading droplet are independent of
the spreading coefficient, the mesoscopic properties, e.g., the length of the
adiabatic film and the thickness of the pancake, do depend on $S_0$. In particular,
the adiabatic film is longer and thinner the larger $S_0$ is.

As we have remarked at the end of Section \ref{macroscopic}, in the
macroscopic picture \cite{deGennes1985} the Young capillary force
$f_Y = S_0 + \sigma_{lg} (1 - \cos(\theta_t))$ per unit length of the contact line is
compensated by the overall viscous force $f_{visc} = \int \left(\eta U/h \right) dx$.
This viscous force can be split into two parts: the viscous force in the macroscopic
wedge which stems from hydrodynamic dissipation in the bulk of the droplet, and the
viscous force in the precursor. By integrating both sides of Eq. (\ref{se}) along the
$x$-axis from the location $x = 0$ of the macroscopic wedge to the tip $x = l_a$
of the precursor film one finds that for arbitrary expressions of the disjoining pressure the viscous force (per length) in the precursor is
exactly equal to $S_0$. This tells that the entire
spreading power $S_0$ is dissipated in the film. This remarkable result due to Hervet
and de Gennes \cite{Hervet,deGennes1985} resolved the long standing paradox that
under complete wetting conditions the spreading of the macroscopic part of the drop
is independent of the value of the spreading power $S_0$.
This is so because this way the driving force of spreading
equals $\sigma_{lg} (1 - \cos \theta_t)$ and it is exactly balanced by
the hydrodynamic dissipation in the bulk so that $S_0$ drops out of the dynamics.

\textbf{(B)} {\it Diffusive films.}
In view of Eq. (\ref{ad}), the adiabatic films may have a noticeable spatial extent
only for very small velocities of the macroscopic wedge. In practice, as soon as the
length of the adiabatic film becomes greater than several micrometers, another
process comes into play. Since the thickness of this film is not constant spatially,
there is a gradient of the disjoining pressure along the film. This causes a non-stationary film -- the so-called ``diffusive'' film -- to develop ahead of the
adiabatic film. When the length $\ell_t$ of such a diffusive film becomes
sufficiently large, the film turns nearly flat so that curvature effects are
negligible. As a result, from Eqs. (\ref{general1}) and (\ref{general2}) one finds
that the profile $h(x,t)$ of the spreading diffusive film solves the differential
equation
\cite{deGennes1984b,Joanny3}
\begin{equation}
\label{spr}
\frac{\partial h}{\partial t} = \frac{\partial }{\partial x}
\left[\left(-\frac{h^3}{3 \eta} \frac{\partial \Pi}{\partial h}\right)
\frac{\partial h}{\partial x}\right]\,.
\end{equation}
This is a non-linear diffusion-type equation with an effective
``diffusion coefficient''
\begin{equation}
\label{cd}
D(h) = -\frac{h^3}{3 \eta} \frac{\partial \Pi}{\partial h}
\end{equation}
which depends on the local height of the mesoscopic film. In the presence of
dispersion forces $\Pi(h) \sim h^{-3}$ and thus $D(h)$ increases as the height
decreases. A straightforward analysis shows \cite{deGennes1984b,Joanny3,Joanny4} that
when the diffusive film becomes much longer than the adiabatic one, the length
$\ell_t$ of such a  film increases "diffusively", i.e., in proportion to the square
root of time:
\begin{equation}
\ell_t \sim \sqrt{D(e(S_0)) t},
\end{equation}
where
$e(S_0)$ is the thickness of the equilibrium pancake shape of the whole drop. Hence,
$D(h = e(S_0))$ is the largest possible value of the diffusion coefficient. In the case
of a van der Waals liquid, $D(e(S_0))$ as given by Eq. (\ref{cd}) is comparable with Stokes diffusion coefficient of a sphere of radius $e(S_0)$, i.e., ca. $10^{-9} \, {\rm m}^2/{\rm s}$.

Thus the phenomena occurring at the macroscopic and mesoscopic scales, where
hydrodynamics is applicable, can be summarized as follows. In case of complete
wetting a macroscopic non-volatile drop spreads very slowly [Eq. (\ref{tanner})] due
to the balance between the hydrodynamic viscous dissipation in the bulk and the
``driving'' Young force (see end of \textbf{(A)} above). It gradually empties into a
mesoscopically thin film, which is formed by the lateral gradient of the disjoining
pressure during the spreading process and serves as a lubricant for the macroscopic
part of the trailing droplet. (This process also removes a singularity in the hydrodynamic dissipation at the three-phase contact line.) The
entire initial spreading power $S_0$ is dissipated by viscous friction in this film.
The mesoscopic film flattens as it spreads. When the entire volume of the
droplet leaks into the film the spreading process stops and an equilibrium
``pancake'' is formed. This description remains valid as long as the thickness
$e(S_0)$ of this final pancake shape is significantly larger than the molecular
size.

\subsection{Non-volatile droplet spreading at microscopic scales:
ultrathin molecular films}

As noted above, the mesoscopic films cannot provide a complete
picture for the
spreading of a non-volatile liquid droplet. The continuum hydrodynamic description
explicitly presumes that the
thickness of the liquid film stays within the mesoscopic range which means that the
thickness $e(S_0)$ of the pancake should be much larger than the molecular scale.
This imposes constraints on the values of the surface tension $\sigma_{lv}$ and of
the spreading power $S_0$. Accordingly, the hydrodynamic description might hold for
low energy substrates, but may break down for intermediate or high energy surfaces,
for which the pancake thickness drops below the size of the liquid molecules.
\footnote[8]{
\label{foot8}
According to the classification of Zisman \cite{Zisman} there are two main types of
solids: (i) hard solids (covalent, ionic, or metallic), and (ii) weak molecular
crystals (bound by van der Waals forces, or in some special cases, by hydrogen
bonds).
Hard solids have ``high energy surfaces'' ($\sigma_{s} \sim 0.5$ to $5\, $ $\rm
N/m$), while molecular solids (and also molecular liquids) have ``low energy
surfaces'' ($\sigma_{s} \sim 0.05 \,$ $\rm N/m$). Here $\sigma_s$
denotes surface energies, i.e., half of the work needed to separate an infinite piece
of material into two half-spaces and take the two emerging solid-vacuum planar interfaces far
from each other.
}
In this context, the challenging experimental observation \cite{cazabat1,beaglehole}
that spreading of non-volatile droplets of squalane or PDMS on silicon wafers
is accompanied by the occurrence of a film of a \textit{microscopic}, not
mesoscopic, thickness called for new experimental approaches and theoretical
concepts.  These will be reviewed in Sections \ref{sec_exper},
\ref{sec_models} and \ref{sec_num} below.

\section{\label{sec_exper} Experimental studies of spreading of microscopic
precursor films}

\subsection{\label{thin}Molecularly thin precursor films in
liquid-on-solid systems}

The experimental analysis of films with thicknesses of the order of only
several molecular diameters became possible with the advent of advanced
experimental techniques such as spatially resolved ellipsometry. Ellipsometry
is an optical method allowing for the measurement of the local thickness of
very thin films which are deposited on substrates
with an optical index of refraction
$n$
different from that of the film.
If the contrast between the indices of refraction is large [such as for silica
or silicon oil ($n = 1.4$ for red light) on a silicon substrate ($n = 3.8$)],
effective film thicknesses as small as $0.1$ \AA~can be measured
\cite{Drevillon,beaglehole}; the lateral resolution remains usually in the
range of micrometers. Here it is important to emphasize that,
in fact, ellipsometric measurements yield ``effective'' thicknesses, which
equal the actual thickness multiplied by the local number density (averaged
across the film at a certain lateral position) divided by the bulk number
density. In other words, the effective thickness is the actual thickness times
the ratio $(n_{loc} - 1)/(n_b - 1)$, where $n_b$ is the bulk index of
refraction for the liquid and $n_{loc}$ is the local index of refraction
(also averaged across the film at a certain lateral position). This local
thickness has the property, that it vanishes, if there is no film, i.e.,
$n_{loc} = 1$. In consequence, effective thicknesses below the molecular
diameter can be observed in ellipsometry which signifies that the film
becomes a diluted, two-dimensional surface gas, rather than a dense fluid.

Employing spatially resolved ellipsometry with modulated polarization, as developed
by Dr\'evillon \textit{et al} \cite{Drevillon} and Beaglehole \cite{beaglehole},
a systematic analysis of spreading speeds of ultrathin precursor films has been
carried out by Heslot \textit{et al} \cite{cazabat1}. They focused on the temporal
evolution of the shapes of rather small drops (with a volume of about $10^{-4} \mu
l$) of non-volatile liquids (squalane and PDMS) spreading (in the complete wetting
regime) on silicon wafers. First, the authors did not observe a ``pancake'' as the final
stage of spreading but rather they detected a gradual transition to a surface gas due
to molecular diffusion on the substrate surface. Second, their analysis revealed a
precursor film of nearly molecular thickness the radial extent of which, $\ell_t$,
increases in time as
\begin{equation}
\label{sqrt}
\ell_t \approx \sqrt{D_{1} t}\,,
\end{equation}
where the subscript ``1'' indicates that, distinct from Eq. (\ref{cd}), $D_1$ is
the ``diffusion coefficient'' related to a molecularly thin precursor film.

Beaglehole \cite{beaglehole} analysed the profiles of the microscopic precursor
of a spreading drop of siloxane oil on glass, fused silica, and freshly cleaved
mica. He
also observed that the mean radial extent $\ell_t$ of the precursor at late stages
follows Eq. (\ref{sqrt}). (Late times mean that they are much larger than the
single
molecule diffusion time, i.e., the time it would take an isolated molecule to
move by diffusion over a distance equal to its diameter, with the lateral extent
of the precursor being orders of magnitude larger than this molecular size, yet
small
enough so that the drop remains large and can act as a particle reservoir.)

\begin{figure}[!htb]
\centering
\includegraphics[width=.8\linewidth]{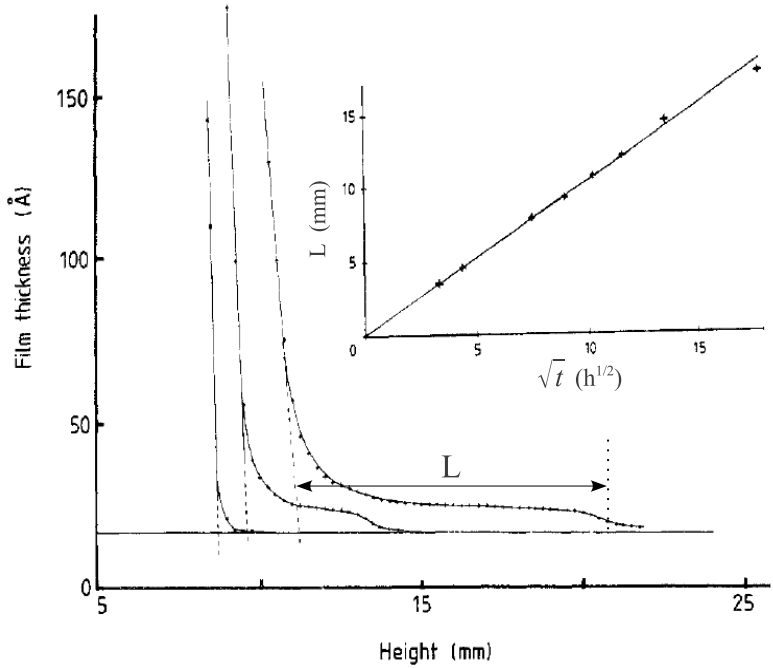}%
\caption{
\label{fig_pdms_films}
Formation of a low-molecular-mass methyl-terminated polydimethylsiloxane
film up a vertical silicon wafer covered with a natural oxide. The curves
from left to right correspond to the ellipsometric effective (see the main
text) thickness profiles (relative to the surface of the silicon wafer)
at 10 min, 10 h, and 56 h, respectively, after the contact of PDMS with
the wafer. The \textit{x} axis is the vertical distance (mm)  measured from
some fixed point within the bulk liquid; \textit{y = 0} defines the silicon
wafer surface and the base line (i.e., the solid horizontal line
at $\approx$ 18 \AA) is the surface of the natural oxide.
The reference position, i.e., the end of the macroscopic meniscus from where the extent $L$ of the film is measured, is indicated in each case by a dashed vertical line. Note that the end of the macroscopic meniscus is also moving the vertical plate upwards, but much more slowly than the film; this latter process has not been studied in detail.
The vertical dotted line at the right indicates the position
of the inflection point in the foot-like region at the advancing edge of
the precursor, which is taken to be the point up to where the film extends.
Inset: film length $L$ as a function of the square root of time (in units
of hours).
[Fig. 1 in F. Heslot, A. M. Cazabat, and N. Fraysse, {\it Diffusion-controlled
wetting films}, J. Phys.: Condens. Matter \textbf{1}, 5793-5798 (1989)
(doi:10.1088/0953-8984/1/33/024). Copyright \copyright 1989, reproduced with
permission from IOP Publishing Ltd.]
}
\end{figure}
Using ellipsometry and $X$-ray reflectometry in a controlled atmosphere of dry
and
filtered $N_2$, Heslot \textit{et al} \cite{cazc} (see also Ref.
\cite{Cazabat1991})
studied the spreading speed and the number density profiles normal
to the substrate of a
molecularly thin precursor film emanating from a macroscopic meniscus in a
capillary rise geometry in which a vertical silicon wafer covered by a natural
oxide
is immersed in a light silicon oil (PDMS). They observed a film  climbing up the
vertical wall the extent of which -- after $56$ hours -- attained macroscopic
values
of ca. $10$ millimeters (see Fig. \ref{fig_pdms_films}). The film number
density varied steplike in the direction normal to the substrate surface. For
the major part of the
film in lateral direction the effective film thickness was nearly constant with
a value of
about 6 \AA. Near the tip of the film its effective thickness decreases
smoothly. Since PDMS
is a worm-like polymer with the size of the monomer of the order of 6 \AA, the
observations have been interpreted such that the major part of the film is a
compact
monolayer of disentangled PDMS molecules lying flat on the solid surface. The
region
near the tip, where the measured effective thickness falls to a value
corresponding to a submonolayer regime, can be viewed as one populated by a
surface gas of PDMS molecules. (Note that this effective thickness can take
values below the size of a PDMS molecule.) The lateral extent $\ell_t$ of the
film (denoted as $L$ in Fig. \ref{fig_pdms_films}) was measured at various
times and it was found to be in agreement with Eq. (\ref{sqrt}) with
$D_{1} \approx 3 \times 10^{-11} ~\mathrm{m}^2/\mathrm{s}$.

\begin{figure}[!htb]
\centering
\includegraphics[width=.6\linewidth]{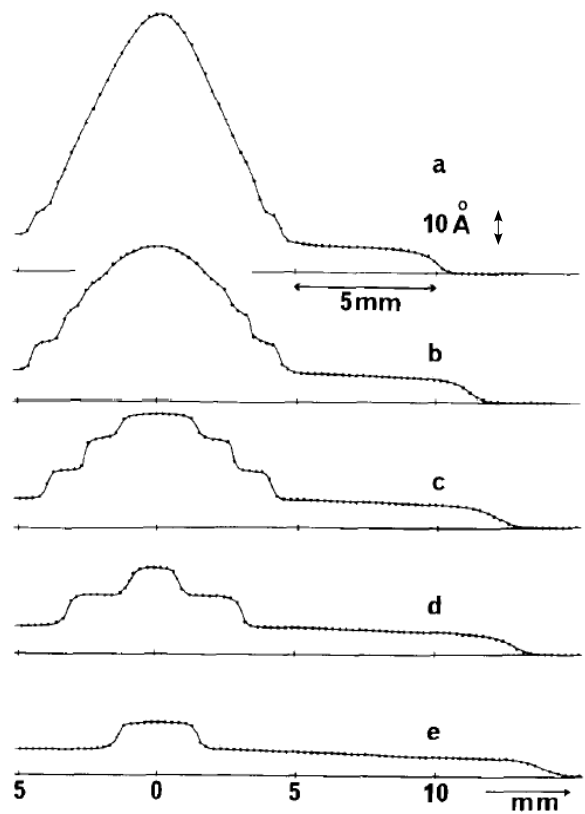}
\caption{
\label{fig_TK_films}
Ellipsometric effective (see the main text) thickness profiles
of ultrathin films as function of
the radial distance from the center of the drop. The data correspond to a
tetrakis(2-ethylhexoxy)-silane (TK) microdroplet with a strongly
layered structure (the TK molecule diameter is 10 \AA). From top to bottom:
72, 83, 120, 144, and 168 h after deposition. [Fig. 2(b) in F. Heslot,
N. Fraysse and A. M. Cazabat, {\it Molecular layering in the spreading of
wetting liquid drops}, Nature \textbf{338}, 640 - 642 (1989)
(doi:10.1038/338640a0). Copyright \copyright 1989, reprinted with permission
from Macmillan Publishers Ltd.]
}
\end{figure}

Significant progress has been made in Ref. \cite{cazabat2} which reports the
striking phenomenon of ``terraced wetting'' (see Fig. \ref{fig_TK_films}).
Using spatially resolved and time-resolved ellipsometry, it was shown that
liquid drops
(PDMS or tetrakis(2-ethylhexoxy)-silane (TK)) spreading on silicon wafers
exhibit a strong, dynamical layering in the vicinity of the solid surface so
that a spreading droplet advances via a series of distinct molecular layers,
such that the  $k$th layer expands proportional to $\sqrt{t}$ with its own
``diffusion coefficient'' $D_k$ ($\dots > D_{k-1} > D_k > D_{k+1} > \dots$).
A similar effect has been observed for spreading of TK in the capillary rise
geometry \cite{qq}. Since the latter terraced spreading occurred for a system
seemingly similar to the one in Ref. \cite{cazc}, for which it was not observed,
this behaviour has been further scrutinized in Ref.\cite{cazabat3}, which
focused on the effects of the surface energy on precursor film spreading.
It was found that ``terraced wetting'' occurs on so-called ``high-energy''
substrates (such as UV-ozone cleaned silicon wafers being used for the
experiment immediately after cleaning). As long as the macroscopic part of the
drop acts as a reservoir, up to five such molecular layer were observed, each
expanding over macroscopic distances proportional to the square root of time.
When the drop starts to run out of material, the top layers were observed to
empty themselves into the lower ones, down to the first one. In contrast,
for low-energy substrates (such as silicon wafers prepared by monolayer
deposition of a fatty acid using the Langmuir-Blodgett technique, see
Ref. \cite{cazabat3}) typically only a single molecularly thin precursor film
has been observed to advance proportional to the square-root of time
\cite{cazabat3}.

The terraced spreading phenomenon has been further studied in
Refs. \cite{cazabat4,cazabat5,valignat} which present a remarkable
catalogue of
many diverse forms which a spreading "terraced"-shaped droplet can adopt under
a variety of conditions for various liquid-on-solid systems. A diversity of
profiles of the film thickness has been observed, with either a gradual (if
any) variation of the thickness as function of the distance from the
macroscopic drop or a rather strong lateral variation. These studies revealed
that the ``terraced wetting'' phenomenon results from an intricate interplay
between the surface energy and friction, and that it occurs for liquids which
are non-volatile both in 3D and in 2D. At room temperature this is the case
for PDMS oils with a viscosity larger than $0.02$ Pa $\times$ s and for the
silicon derivative TK, and at temperatures below $5^\circ $ C for squalane.

Perfectly terraced droplets with terraces possessing flat tops and sharp edges
have been observed by Daillant \textit{et al} \cite{daillant2,Bardon1999}
for the liquid crystal $8CB$ spreading on a (400) silicon wafer and by
Betel{\'u} \textit{et al} \cite{betelu} for the liquid crystal $\overline{7}S5$
spreading on an oxide covered (100) Si wafer. Using ellipsometry, for
perfluoropolyalkylether (PFPE) films on carbon surfaces Ma \textit{et al}
\cite{ma} have also observed a complex terraced spreading with an increase of
the size of the individual layers proportional to the square root of time,
while  by using surface plasmon resonance Lucht and Bahr \cite{lucht} have
revealed a terraced structure on gold for the spreading of a liquid-crystalline
smectic-A droplet. Similar observations (i.e., terraced spreading or single
monolayer spreading of the precursor with an expansion proportional to
$\sqrt{t}$) have been further reported for a variety of complex
liquids, such as liquid crystals or alkanes, in contact with bare silicone
wafers or ones which are grafted with polymer brushes
\cite{Ramdane1998,Lazar2005,Bardon1999,Xu2000}. Naturally, in such systems the
details of the spreading behavior, such as the prefactor $D_1$, differ and the
structure of the various spreading monolayers is significantly more
complicated.

\begin{figure}[!htb]
\centering
\includegraphics[width= .7\linewidth]
{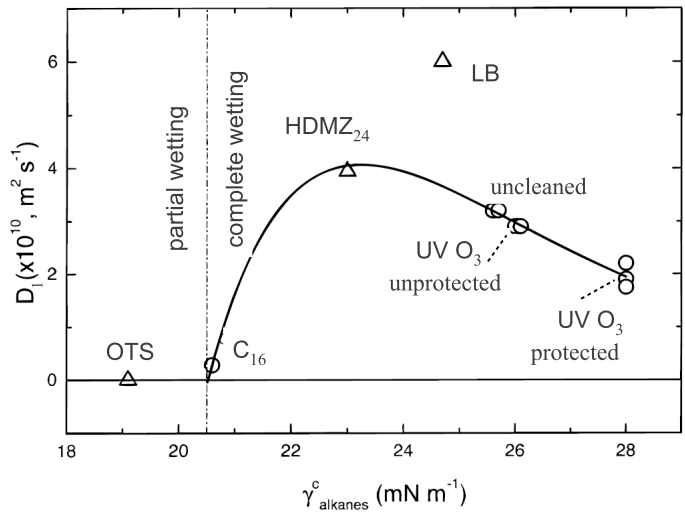}
\caption{
\label{fig_non_monotonic_prefactor_part2}
Film diffusion coefficient $D_1$ for PDMS 20 (molecular mass: 2000;
viscosity:
20 cP; surface tension: $20.6 mN m^{-1}$) versus critical
surface energy ($\gamma^c_\mathrm{alkanes}$) of the substrates
evaluated for the alkane series. The full line corresponds to the best
fit of the data with a theoretical model (Eq. (12) in  Ref. \cite{voue}).
The fit includes only the open \textit{circles}, see Ref. \cite{voue}
for details; note that the triangle corresponding to the case labeled by
HDMZ$_{24}$ falls nonetheless on this line. This is, however, not surprising
because the surface treatment used to produce this substrate (see
below) does not make the HDMZ$_{24}$ surface too different from
those of the "original" wafers, although it lowers the surface energy due to
a trimethyl layer grafted on the native silicon oxide.
The data, represented by open circles and triangles, correspond to
chemically
different  and differently treated substrates: ``uncleaned''
denotes bare wafers used without any cleaning procedure;
``UV O$_3$  unprotected'' wafers
were UV-ozone cleaned wafers (1 h under a dry O$_2$ flow plus
20 min under an O$_2$ flow saturated with H$_2$O);
``UV O$_3$ protected'' wafers were wafers cleaned according
to the same procedure but protected after cleaning by aluminum foils and
stored under nitrogen in the presence of silica gel to prevent a fast
contamination of the reaction sites;
``C$_{16}$'' wafers are surfaces obtained by chemical grafting a 3.4 nm
thick layer of hexadecyltrichlorosilane;
``HDMZ$_{24}$'' surfaces are obtained by exposing an UV-ozone cleaned wafer to
hexamethyldisilazane vapour during 24 h at room temperature (this
procedure results in grafting a 0.4 nm thick trimethyl layer)
and  finally,
``OTS'' wafers are obtained by chemical grafting
octadecyltrichlorosilane on a UV-ozone cleaned wafer (this procedure
results
in a 3.2 nm thick layer). LB denotes a Langmuir-Blodgett wafer which
has a behavior completely different from the others
showing that the associated $\omega$-tricosenoic acid layers drastically
modify the chemistry of the surface.
[Fig. 9 in {\it Dynamics of Spreading of
Liquid Microdroplets on Substrates of Increasing Surface Energies},
M. Vou{\'e}, M. P. Valignat, G. Oshanin, A. M. Cazabat, and J. De Coninck,
Langmuir \textbf{14}, 5951 - 5958 (1998) (doi: 10.1021/la9714115).
Copyright \copyright 1998, reprinted with permission from the
American Chemical Society.]
}
\end{figure}

\begin{figure}[!htb]
\centering
\includegraphics[width= .7 \linewidth]
{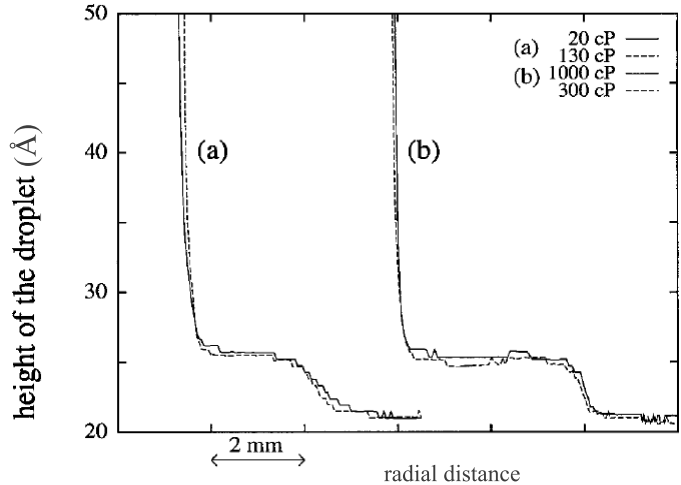}
\caption{
\label{fig_non_monotonic_prefactor_part1}
Experimentally determined shapes of several droplets of PDMS
of different molecular mass $M$ and polydispersity index $I_p$
(as usual defined as the ratio between the
\textit{weight averaged} molecular weight and the
\textit{number averaged} molecular weight in a given polymer sample)
spreading
on one and the same silicon wafer at relative humidity 98\%.
The profiles are recorded at:
(a) 25 min after deposition [solid line: $M$ = 2000 g/mol, $I_p = ­ 1.7$;
dashed line: $M$ = 7300 g/mol, $I_p = ­ 1.10$] and (b) after 40 min [solid line:
$M$ = 28400 g/mol, $I_p = ­ 1.29$; dashed line: $M$ = 13000 g/mol,
$I_p = ­ 1.15$]. The legend indicates the corresponding
viscosities, in centipoise, of the PDMS oils. Note that the spreading of the
macroscopic part of the drop (the macroscopic edge is indicated by the almost
vertical lines) is significant, but also seemingly independent of the oil
type. The reference plane (silica plus adsorbed water) is at 21 \AA.
[Fig. 2 in {\it Molecular Weight Dependence of
Spreading Rates of Ultrathin Polymeric Films}, M. P. Valignat, G. Oshanin,
S. Villette, A. M. Cazabat, and M. Moreau, Phys. Rev. Lett. \textbf{80},
5377 - 5380 (1998) (doi: 10.1103/PhysRevLett.80.5377).
Copyright \copyright 1998, reprinted with permission from the American
Physical Society.]
}
\end{figure}

The influence of the surface energy has been analyzed in detail  by
Vou{\'e} \textit{et al} \cite{voue} who examined the dependence of $D_1$
for the precursor monolayer next to the substrate on the surface energy
characterized by the so-called critical surface energy $\gamma_c$ (obtained
for alkane series as the abscissa of the point at which the extrapolated
linear relation $\cos \theta_{\infty}$ vs $\sigma_{lv}$ -- the so-called
Zisman plot -- intercepts the line $\cos \theta_{\infty} = 1$). By using
spatially
resolved ellipsometry, these studies showed that $D_1$ is equal to zero up to
a certain threshold value of the surface energy followed by a non-monotonic
dependence on the surface energy, exhibiting a pronounced maximum for
intermediate surface energies (see
Figs. \ref{fig_non_monotonic_prefactor_part2} and
\ref{fig_non_monotonic_prefactor_part1} and Sec. \ref{sec_models}).

The effect of the droplet size on the dynamics of such ultrathin films has
been studied by Leiderer \textit{et al} \cite{Leiderer1992}. Using the method
of optically excited surface plasmon resonance, they determined the
density profiles of submonolayer films during spreading of picoliter-volumes of
PDMS on a silver substrate. The results showed that the spreading of the film
is described by the $\sqrt{t}$ dependence on time which may become significantly
slower if the drop is too small to act as a reservoir for the emanating
precursor film.

By using scanning micro-ellipsometry and scanning X-ray photoemission
spectroscopy, Novotny investigated the spreading behaviour of droplets of
perfluoropropylene polymers which are widely used in magnetic recording
applications \cite{novotny}. He observed microscopic precursor films with a
sharp edge and spreading in accordance with Eq. (\ref{sqrt}). He systematically
analysed the dependence of $D_1$ on the molecular weight $M$ and concluded that it can
be described effectively by an algebraic law $D_1 \sim M^{-\alpha}$ with
$\alpha \approx 1.7$. Fraysse \textit{et al} \cite{fraysse} (see also
Ref. \cite{valignat}) studied the spreading speeds of low molecular weight PDMS
precursor films on bare oxidized silicon wafers, used without cleaning, or on
wafers bearing loosely grafted trimethyl layers. Their analysis confirmed that
the radial extent of the film grows proportionally to the square root of
time [Eq. (\ref{sqrt})] and it showed that $D_1$ is inversely proportional
to the bulk viscosity $\eta(M)$ of the PDMS liquid. Since for this range of
molecular weights the best fit of the PDMS
viscosity is given by $\eta(M) \sim M^{1.7}$ \cite{fraysse}, the
observations made in Ref. \cite{fraysse} appear to agree rather well with the
experimental results obtained by Novotny \cite{novotny}. Further on, O'Connor
\textit{et al} \cite{oconnor1,oconnor2} and Min \textit{et al} \cite{min}
studied the spreading behaviour of PFPE (perfluorinated
polyether, terminated with hydroxyl groups), AM2001 (pyperonyl groups), and
Ztetraol (propylene glycol ether groups) on silica surfaces as a function of
the end group functionality, molecular weight, temperature, and humidity.
The square root of time law for the growth of the radial extent of the film
has been observed for all liquid-on-solid systems studied in
Refs.\cite{oconnor1,oconnor2,min}. But the dependence of $D_1$ on the
molecular weight $M$ turned out to be somewhat weaker, i.e., the observed
values of the exponent $\alpha$ seem to be smaller than $1.7$. (However, one
should keep in mind the merely effective character of this power law.) For
example, ellipsometry measurements of ultrathin (i.e., less than 10~\AA~thick)
PFPE precursor film spreading on cleaned silicon wafers with a
native oxide layer rendered $\alpha \approx 0.5$ \cite{min}, which does not
agree with the results of Novotny \cite{novotny} and Fraysse
\textit{et al} \cite{fraysse}.

This issue has been revisited by Valignat \textit{et al} \cite{val1,val2},
who studied the molecular-weight dependence of $D_1$ in Eq. (\ref{sqrt})
for molecularly thin precursor films of PDMS spreading on hydrophilic oxidized
silicon wafers. The wafers were cleaned by oxygen flow under UV illumination
and then placed in a measurement chamber for $24$ hours. The atmosphere
therein contained controlled relative humidity (RH), in order to obtain on
the wafer an equilibrated submonolayer of water with controlled coverage,
depending on RH.
This is motivated by the fact that the substrates used in these experimental
studies are chemically heterogeneous and contain different types of surface
sites. There are low energy siloxane bridges and high energy, chemically
active silanol sites which may form a hydrogen bond with any of the monomers
of a PDMS molecule and thus for the latter represent preferred spots of
attachment to the surface. The water molecules have a strong affinity to the
silanol sites and effectively screen them, forming molecularly thin islands
of water. Hence, exposure to an atmosphere with controlled RH prior to
deposition of a PDMS droplet allows one to modify the frictional properties
of the substrate by changing the chemical composition of the surface.
It was realized that at high substrate friction $D_1$ is controlled by
mesoscopic effects, i.e., by the bulk viscosity $\eta(M)$, and thus as noted
above the exponent $\alpha$ can be substantially larger than $1$. At moderate
RH, i.e., at moderate friction, one observes $D_1 \sim 1/M$ while for very
high RH, i.e., low friction, $D_1$  (and thus the spreading of
the precursor) becomes independent of $M$ \cite{val1,val2} (see
Fig. \ref{fig_non_monotonic_prefactor_part1}, which corresponds
to RH = 98 \% and shows no significant difference between the full and dashed
lines, and Sec. \ref{sec_models}).

Furthermore, Villette \textit{et al} \cite{Tiberg1} focused specifically on
the role of water on the spreading of molecular films of PDMS, PDMS with
hydroxyl ends (PDMS-OH) and TK on oxidized silicon wafers. Their analysis
revealed the dependence of $D_{1}$ on RH. In particular, it was realized that
for PDMS and PDMS-OH the ``diffusion'' coefficient $D_{1}$ shows a distinctly
different linear variation with the relative humidity (and, respectively,
with the substrate coverage by water) for RH less than or larger than $70 \%$.
For moderate RH, $D_{1}(RH)$ has a relatively small slope, which at
RH $\approx 70\%$
crosses over to a very steep linear dependence on RH. Overall, within the range
of $20\%$ to $90\%$ RH, $D_1$ varies by more than two orders of magnitude,
attaining the values observed for mesoscopic precursor films. This agrees well
with the observations made by Valignat \textit{et al} \cite{val1,val2}.
Such a remarkable enhancement of $D_1$ has been explained by the fact that at
this value of RH the patches of water on the substrate start to overlap and
form a tortuous connected structure on which the spreading of molecules
encounters a very low friction. Surprisingly, the dependence of $D_{1}$ on RH
for TK spreading on hydrophilic substrates appears to be more complicated.
Similarly to PDMS and PDMS-OH, $D_1$ gradually increases up to RH $\approx 70\%$,
followed by an abrupt growth. However, at RH $\approx 80\%$ $D_{1}$ passes
through a maximum and then starts to decrease. At this latter RH
one observes signs of a ``dewetting transition'', in the sense that the previously
distinct first, second and third layers in the terraced shape do no longer
occur and only the first layer spreads, while the rest of the drop does not. For RH larger than $85\%$ the behaviour turns into that for partial wetting: while a monolayer precursor film spreads ahead of the drop, the macroscopic part of the drop attains a non-zero equilibrium contact angle.

Within the complete wetting regime, the $\sqrt{t}$ time dependence of a
spreading monolayer has also been observed in the case of PDMS on highly
polished silicone wafers coated with a self-assembled monolayer of octadecyl-trichlorosilane (OTS)\cite{Steiner2001}. In this study an interferometric
video microscopy technique was employed. See also Ref. \cite{Kavehpour2003},
in which interferometric microscopy techniques have been employed in order to
study the spreading of small drops of very viscous, low-volatile liquids on
planar smooth substrates.) The authors have been able to accurately
reconstruct the shape of the drop during spreading; from the shape, the volume
of liquid in the drop was computed as a function of the time $t$ elapsed
since the
beginning of spreading. Since under the corresponding experimental conditions
the liquid used is practically non-volatile, any change (i.e., decrease) in
the volume $V(t)$ enclosed by the drop shape has been attributed to liquid
being transferred to a precursor film. Therefore, although such films have
not been directly imaged and investigated, their existence and the fact that
their linear extent grows in time proportionally to $\sqrt{t}$ could be clearly
inferred from the significant changes in $V(t)$ (see Fig. 8 in
Ref. \cite{Steiner2001}).

Finally, we mention several other situations in which spreading ultrathin
precursors have been observed for partially wetting liquids. Tiberg and
Cazabat \cite{Tiberg2} have studied the spreading behaviour of non-ionic
trisiloxane oligo(ethylene) oxide surfactants on high and low energy surfaces.
These surfactant molecules have a peculiar  ``hammer''-like structure and
consist of a compact hydrophobic trisiloxane group sitting as a ``hammer head''
on an ethylene oxide, which forms a $35$~\AA~long ``dangling handle''.
It was shown that the spreading of drops of these liquids, if
at all, occurs only via an autophobic ultrathin precursor film emanating from
the drop (``autophobic'' in the sense that the drop ``does not wet its own
film'' but rather remains in a partial wetting state on top of the
precursor-covered substrate). On low energy surfaces these surfactants
self-assemble over the solid into well defined bilayers the radial extent of
which grows proportional to $\sqrt{t}$, and after a few days in humid
atmosphere they cover several square centimeters.
The thickness of the precursor is virtually constant ($54 \pm 2$ \AA)
during the whole spreading process. On high energy surfaces, spreading again
occurs via an autophobic precursor, the linear extension of which also grows
proportional to $\sqrt{t}$. But the shape of the  vertical cross section of the precursor is now strikingly different having a cone-like shape
with a very wide opening angle and a rounded tip rather than a compact organized structure.
This aspect has been further studied in Ref. \cite{Tiberg3}, which focused
on spreading of various hammer-shaped and linear-shaped surfactants on
low, intermediate and high energy surfaces. It was confirmed that spreading
of a non-wettable liquid of surfactants proceeds via a ultrathin precursor
film with $\ell_t \sim \sqrt{t}$ and only the shape of the vertical cross
section of the precursor film depends on the particular choice of the
surfactant and of the solid substrate.

\subsection{\label{metal_metal_exper}Ultrathin precursor films in
metal-on-metal systems}

As discussed above, most of the studies of precursor films have been focused
on the spreading of liquids on solid substrates. A decade ago, it has been
pointed out that microscopically thin films extending over macroscopic
distances also occur for wetting in solid-on-solid systems in the so-called
Stranski-Krastanov state (crystallites coexisting with a film, see Refs.
\cite{Prevot2000,Humfeld2004} and references therein). A particular
example, which has been the subject of a number of also recent studies
\cite{Moon2001,Moon_la2004,Moon_ss2004,Monchoux2005,Monchoux2006,Wynblat2007},
is that of metal films (Pb, Bi or Pb-Bi alloys) on a metal substrate
(mono-crystalline Cu(111), Cu(100) or polycrystalline Cu). As we shall
discuss below, these studies have provided evidence for submonolayer
precursor films exhibiting the same spreading dynamics (linear extent
growing in time as $\sqrt{t}$) as in the cases of liquid spreading.
They allowed a clear
identification of surface diffusion through the film as the mechanism of
precursor formation and spreading.

At not too elevated temperatures (among these metals, at ca. 544.7 K Bi has
the lowest bulk melting point), such systems intrinsically have a very
low vapour pressure. The choice of Pb, Bi or Pb-Bi alloys to be deposited
on Cu(111) (triangular lattice \cite{Moon2001,Moon_ss2004}) or Cu(100)
(rectangular lattice \cite{Prevot2000}) is motivated by the practically
negligible bulk solubility of Pb or Bi in Cu. This prevents loss of material
via diffusion into the substrate. But there is a good bulk solubility of Bi
in Pb, which allows one to prepare alloys over a wide range of
concentrations. Moreover, Bi and Pb atoms are very similar in size, which
makes the theoretical interpretation of the results easier. In all cases,
physical vapour deposition from high-purity metal sources under ultra-high
vacuum was used to create thick (i.e., hundreds of nanometers) overlayers
of metal
under very well controlled atmospheric conditions. During the preparation
of the experiment several cycles of sputtering have been employed to
ensure a contaminant-free surface. The investigation of the emerging
precursors has also been performed under high-vacuum conditions. The
experiment encompassed a combination of scanning Auger electron
spectroscopy (SAM) and Scanning Tunneling Microscopy (STM) or Scanning
Electron Microscopy (SEM). The former method serves to determine the
atomic composition; it has a spatial resolution below one micron and
requires ca. 5 min for scanning \cite{Moon2001,Moon_ss2004,Moon_la2004}.
With the latter methods one can obtain roughly the spatial extent of the
film or
the shape (i.e., the contact angle) of the Pb or Pb-Bi clusters on the
surface.

In the investigations of Pb and of Pb-Bi alloys on Cu(111)
\cite{Moon2001,Moon_ss2004}, an initially thick (i.e., hundreds of nanometers)
overlayer of metal was gradually heated up to slightly above the melting
point of the material (for Pb this is ca. 600 K, and for alloys ca.
538 - 568 K), which led to dewetting and the formation of drops
coexisting with a thin film; this configuration was quenched.
\footnote[9]{
\label{foot9}
The authors interpreted these observations as ``pseudo-partial'' wetting.
This confusing nomenclature was clarified later\cite{Bonn2005}.
}
Even in the solid state the droplets are observed to maintain a quasi-spherical
shape with just a flattened top;  the ``dome'' rather than
``pyramid'' shape could be the result of solidification under a rapid temperature
quench.  This observation allows one to assign an approximate contact angle
even for the solidified particles. As reported in
Refs. \cite{Moon_ss2004,Moon_la2004}, for temperatures $T$ in the range
$360 \, \mathrm{K} \lesssim T \lesssim 610 \, \mathrm{K}$, these contact angles are
ca. $40^\circ - 45^\circ$, indicating a partial-wetting situation, with a
very weak dependence on temperature and composition. Upon annealing
at a desired temperature (note that by varying the temperature one can
also control the timescale of the dynamics), by waiting for thermal
equilibration and by sputtering off the film (which also ensures a final
decontamination), the authors were able to investigate the formation and the
spreading of precursor films in contact with liquid drops or solid particles
\cite{Moon2001,Moon_ss2004,Moon_la2004}.  For Bi, a different setup was used,
in that the film deposition has been performed on a partially masked
substrate, while the remaining steps of the procedure, i.e., annealing and
sputtering, were the same. This allowed the study of a precursor formation
from a straight contact line \cite{Moon_ss2004,Moon_la2004}.
Therefore, both straight and circular configurations of the film have been
investigated, in addition to the various possible combinations of
liquid or solid drops or films.

\begin{figure}[!htb]
\centering
\includegraphics[width=.75 \linewidth]{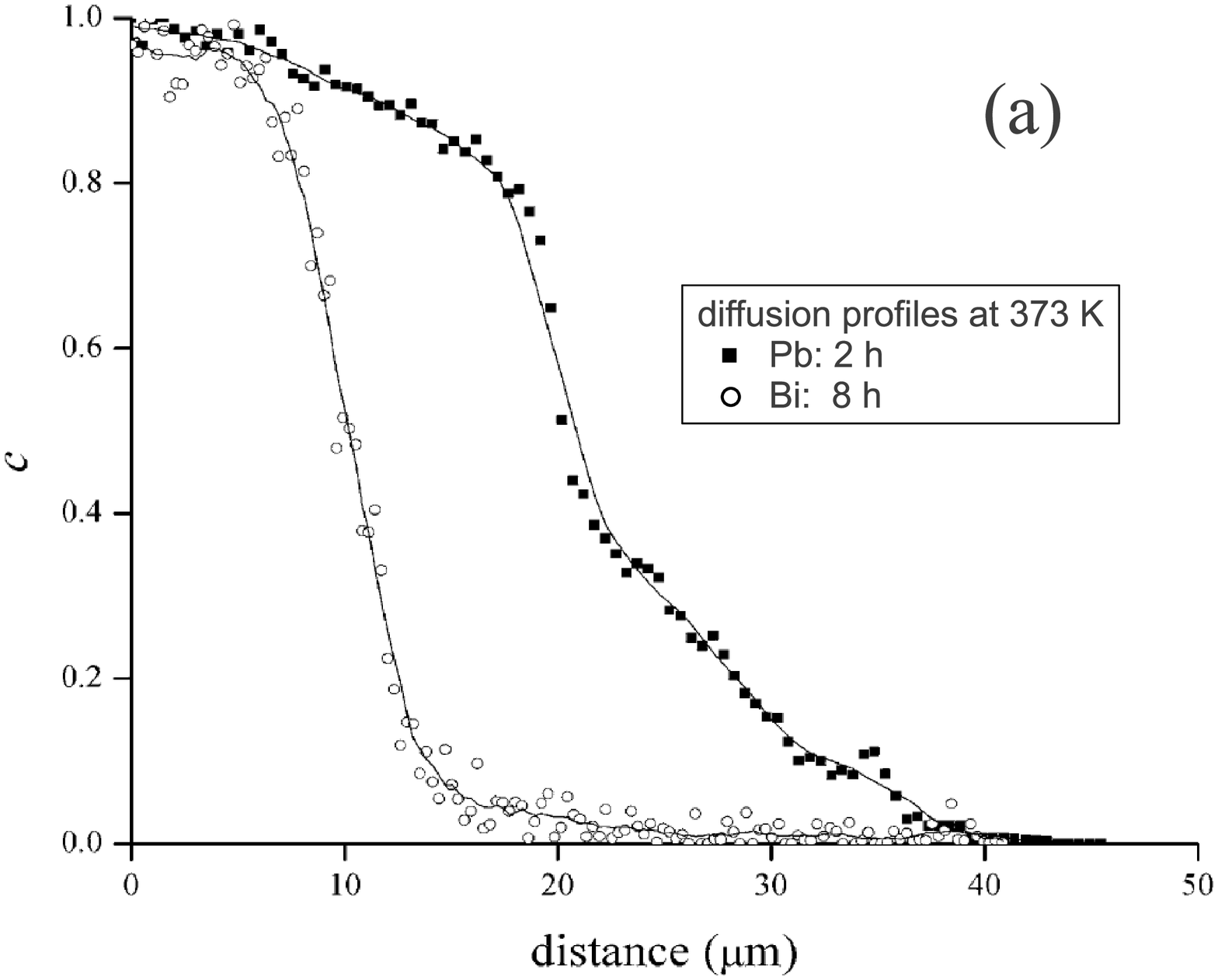}
\includegraphics[width=.75 \linewidth]{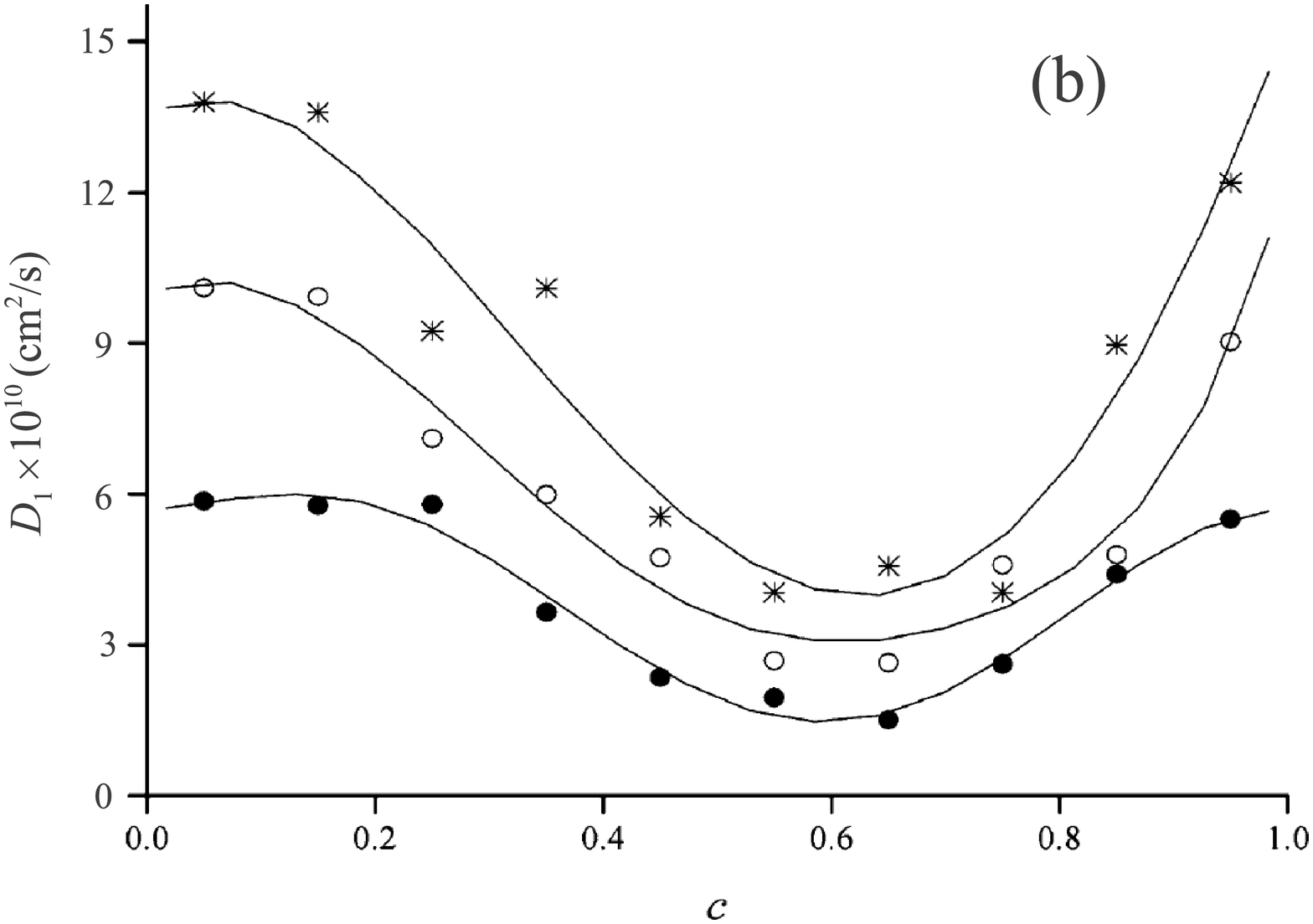}%
\caption{
\label{Pb_Bi_exper}
\textbf{(a)} Profiles of the coverage $c$ for Bi (at $t = 8$ h) and for Pb
(at $t = 2$ h) on Cu(111) at T = 373 K.
[Fig. 5 in {\it Diffusion kinetics of {\rm Bi} and {\rm Pb - Bi} monolayer
precursing films on {\rm Cu(111)}}, J. Moon, P. Wynblatt, S. Garoff, and R.
Suter, Surface Science \textbf{559}, 149-157 (2004)
(doi:10.1016/j.susc.2004.04.018). Copyright \copyright 2004, reprinted with
permission from Elsevier.]
\textbf{(b)} Coverage dependence of the diffusion coefficient of Pb on Cu(111)
at temperatures 353 K ($\bullet$), 363 K ($\circ$) and 373 K ($\ast$).
[Fig. 9b in {\it Pseudopartial Wetting and Precursor Film Growth in Immiscible
Metal Systems}, J. Moon, S. Garoff, P. Wynblatt, and R. Suter,
Langmuir \textbf{20}, 402-408 (2004) (doi:10.1021/la030323j).
Copyright \copyright 2004, reprinted with permission from the American
Chemical Society.]
For further details see the main text.
}
\end{figure}

The results of these experiments revealed that in all cases a precursor
film of submonolayer thickness emanates from the drop, and the linear
extent $\ell_t$ of this film obeys the relation given in Eq. (\ref{sqrt})
with the prefactor $D_1$ depending on geometry, the metal, the composition
of alloys, and temperature. This is very similar to the features observed
in the experiments dealing with the spreading of liquids. Somewhat
controversially, changes in the temperature dependence of $D_1$ have been
interpreted as to indicate a change in the state of the film between a
disordered (fluid) and a partially ordered (solid) structure
\cite{Moon_la2004}. Similar changes would occur due to, e.g., a change in
the dynamics induced by a partial surface alloying.
However, in the case of pure Pb or Bi, the coverage $c$ (i.e.,
an analogue of the effective film thickness $h$ discussed in
Subsec. \ref{thin}), where $c$ is defined such that it
takes the value of one (corresponding to a compactly filled monolayer) at
the edge of the drop or at the straight contact line, depends strongly on
the radial distance $x$ from the edge of the drop or from the straight
contact line [see Fig. \ref{Pb_Bi_exper}(a)]. In all these cases,
dividing the spatial coordinate by $\sqrt{D_0 t}$ ($D_0$ is
the corresponding diffusion coefficient of Pb or Bi atoms on the Cu surface)
leads to a collapse of the profiles taken at different times $t$ onto
a single master curve $\Theta(\lambda = x/\sqrt{D_0 t};\,T,\mathrm{~metal})$
the shape of which depends on the temperature $T$ and on the materials
involved. This scaling is compatible with a surface diffusion mechanism.
\footnote[1]{
\label{foot10}
Although Refs.\cite{Moon_ss2004,Moon_la2004} mention studies of Pb
films at temperatures of $600$ K and above, for which the film would be in
a disordered state and the drop in the liquid state, no coverage profiles at
these temperatures are presented.}
The master curve $\Theta(\lambda;\dots)$ does not necessarily have the shape
of an error-function, as it would be the case for a simple one-dimensional
diffusion process with a fixed concentration source at the origin,
but in many cases it exhibits steep drops within certain ranges of
coverage \cite{Moon_ss2004,Moon_la2004}. This indicates a coverage-dependent
diffusion coefficient, which might be caused  by certain surface alloying
processes as discussed in detail in
Refs. \cite{Prevot2000,Humfeld2004,Moon_ss2004,Moon_la2004}, but also by
attractive interactions between the adsorbed atoms (see Sec. \ref{sec_models}).
Via a Boltzmann-Matano type analysis \cite{Matano1933}, from such
concentration profiles the coverage dependence of an effective diffusion
coefficient $D(c)$ has been obtained. For both Pb and Bi, the resulting
diffusion coefficients $D(c)$ have a complex structure, with one \cite{Moon2001}
or even two minima \cite{Moon_la2004} at intermediate coverages and maxima at
high coverages [see Fig.  \ref{Pb_Bi_exper}(b)]. The positions of the minima
are compatible with known surface alloying transitions for Pb or Bi on Cu(111).
Similar findings have been reported for films of Pb or Bi on Cu(100)
\cite{Monchoux2006}. The results for binary alloys are similar, in the sense
that a Pb-Bi film spreads on the Cu(111) substrate, but the rate of spreading
seems to be dictated by the slower component and the composition of the alloy
varies significantly with the distance from the edge of the drop.

Finally, we note that an interesting extension of this system was studied in
Ref. \cite{Monchoux2005}, involving two macroscopic films of Pb and Bi,
respectively, which are deposited such that they are separated by a small area
of the substrate and give rise to precursor films that ``collide''. By using
SAM, the composition profiles of both Pb and Bi along the direction of mutual
approach have been determined. This allows a study of two-dimensional surface
interdiffusion of Pb and Bi.

We conclude this section by emphasizing several important features extracted
from the experimental analyses of spreading of molecularly thin precursor
films.

\begin{enumerate}[(1)]
 \item Precursor films are omnipresent, even if the liquid droplets only
partially wet the solid surface or, in the extreme case, if the drops are
actually solidified clusters.

 \item Precursor films spread from the drop as a result of the competition
between the gain in entropy and the balance of the attractive interactions
between the fluid molecules and the substrate and the attractive interactions
among the fluid molecules. Condensation out of the vapour may, in principle,
occur in addition but it is not \textit{condicio sine qua non} for the
formation of precursor films.

 \item Precursor films act as a lubricant for the drops in that the
macroscopic part of the drops spreads on a "prewetted"
substrate.

 \item At sufficiently large times (but with the drop still being able to act
as a reservoir for the precursor film), the radial extent of the film increases
proportional to the square root of time. This dependence holds very generally.
Only the prefactor $D_1$ [Eq. (\ref{sqrt})] depends on the particular
features of the liquid-on-solid system under study.

 \item One encounters different types of thickness profiles.
(For submonolayer films the coverage is translated into an
\textit{effective} thickness, see the discussion of Fig. \ref{fig_pdms_films}).
In some instances, molecular-sized precursor films are compact and the
thickness profile is laterally constant. In other cases, the thickness along
the film varies considerably.

\end{enumerate}

\noindent These observations are the basis for formulating a theoretical
understanding of the spreading of precursor films.

\section{\label{sec_models}Models for the dynamics of spreading of
microscopic precursors}

The theoretical analyses of the physical mechanisms underlying the seemingly
universal $\sqrt{t}$-law and the ``terraced wetting'' phenomenon have followed
three different lines of thought.

\begin{figure}[!htb]
\centering
(a)\hspace*{.4\linewidth}(b)\hfill\\
\includegraphics[width=.52 \linewidth]{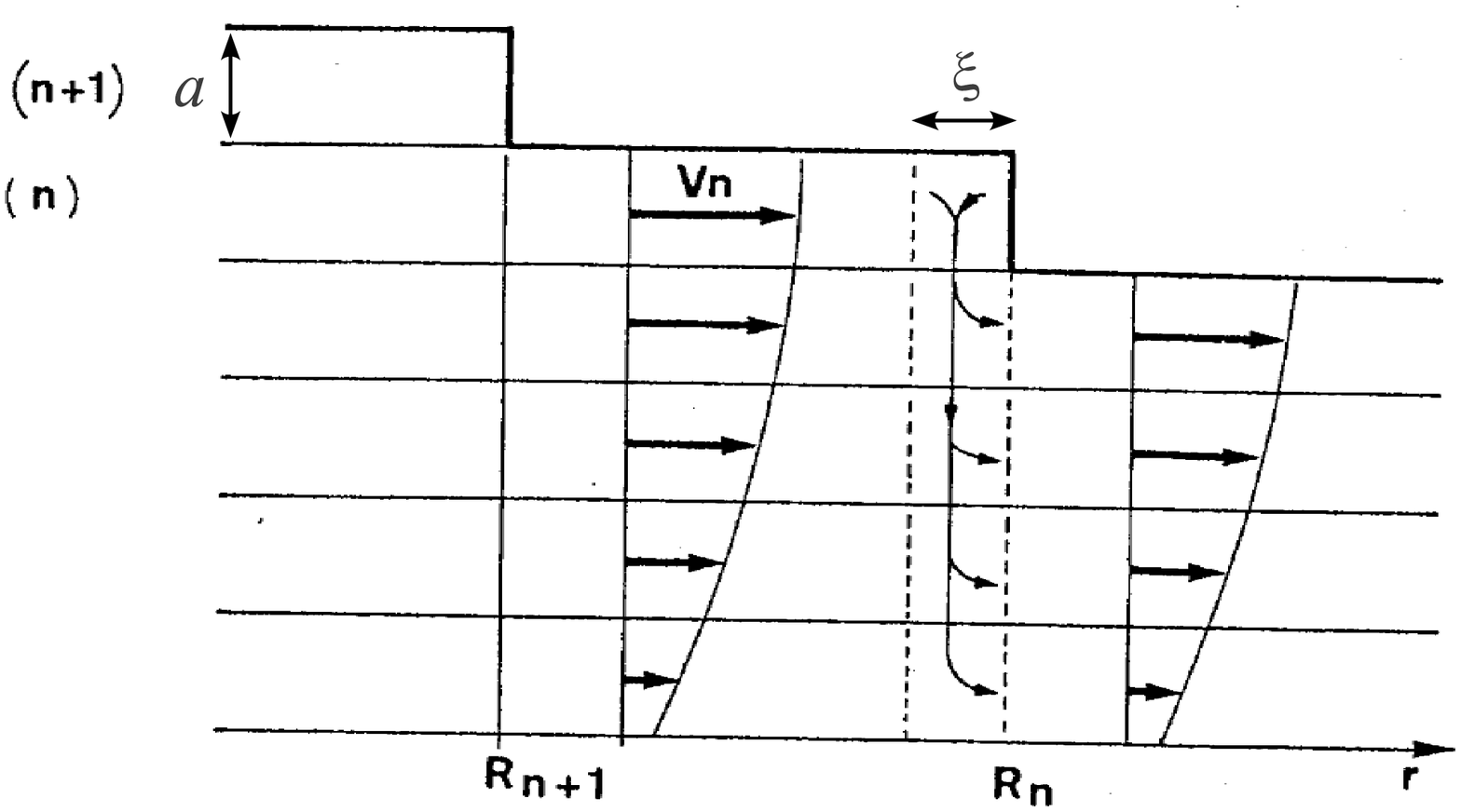}%
\hspace*{.05 \linewidth}%
\includegraphics[width=.43 \linewidth]{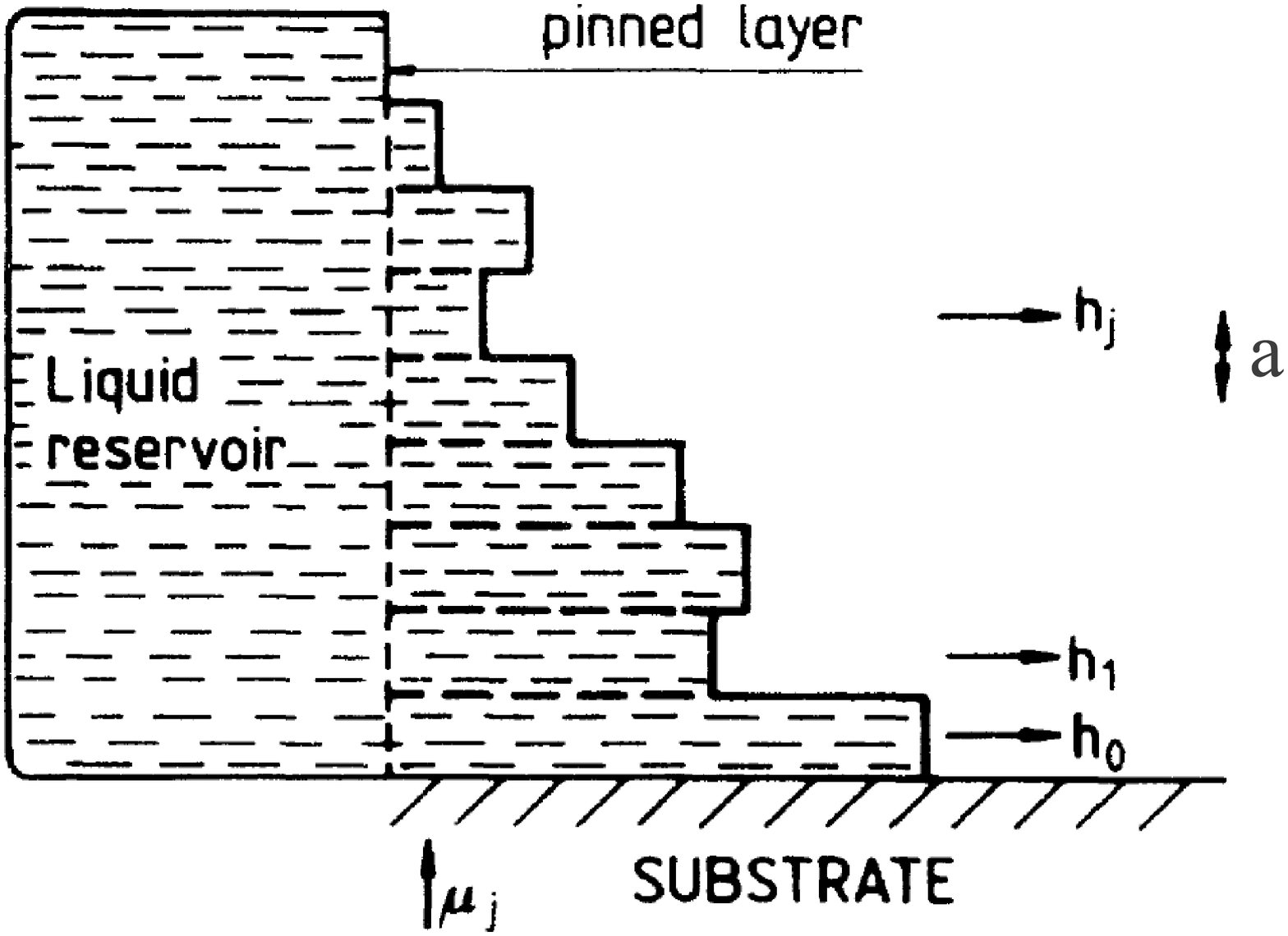}%
\caption{
\label{layered_models}
{\bf (a)} Schematic drawing of the layered structure of a drop as considered in Ref. \cite{pgg2}. In the main text the radial extent $R_n$ of the $n$th
layer is denoted as $l_n$; $a$ is the thickness of a layer and $\xi$ denotes
the extent of the permeation zone (i.e., where, as indicated, flow occurs
along the vertical direction) located at the advancing edge of the topmost
layer at a certain $R_n$ and marked by the vertical dashed lines. The
velocity field in the $n$th layer is denoted as ${\rm V}_n$.
[Fig. 1 in {\it Spreading of a stratified incompressible droplet}, P.G. de
Gennes and A. M. Cazabat, C. R. Acad. Sci. Paris II \textbf{310}, 1601-1606
(1990). Copyright \copyright 1990, reprinted with permission from the
Acad{\'e}mie des Sciences -- Institut de France.]
{\bf (b)} Schematic drawing of the horizontal solid-on-solid model considered
in Refs. \cite{joela,joelb,joelc}. Here the lateral extent of the $j$th layer
is denoted as $h_j$ (as opposed to $l_j$ in the main text) and $\mu_j$ (the
analogue of $W_n$ as used in Ref. \cite{pgg2}) denotes the interaction with
the substrate of the form $\mu_j = \mu_0 f(j)$, where $f(j)$ is a decreasing
function of the height $j$ above the substrate and $\mu_0$ is proportional
to the Hamaker constant for the system under consideration.
At a certain height $j = L$ above the substrate
the lateral extent of the corresponding layer is fixed (pinned
layer) which mimics the macroscopic wedge which is taken to not evolve with
time. Note that for some layers $h_{j+1} > h_j$ as a result of
thermal fluctuations.
[Fig. 1 in {\it Dynamics of a microscopic droplet on a solid surface: Theory
and simulation}, J. Heini{\"o}, K. Kaski, and D. B. Abraham,
Phys. Rev. B \textbf{45}, 4409 - 4416 (1992)
(doi: 10.1103/PhysRevB.45.4409). Copyright \copyright 1992, reprinted with
permission from the American Physical Society.]
}
\end{figure}

De Gennes and Cazabat \cite{pgg2} proposed an analytical description of the
``terraced wetting'' phenomenon, in which the liquid drop on a solid surface
was considered as a completely layered structure, the $n$th layer being a
quasi two-dimensional, incompressible fluid  of molecular thickness $a$
and with macroscopic radial extent $R_n$ [see Fig. \ref{layered_models}(a)].
The interaction energy of a molecule in the $n$th layer with the solid
substrate was taken to be of the general form of a negative, increasing
(towards zero) function $W_n$ of the distance $n \times a$ from the substrate.

Within this model de Gennes and Cazabat \cite{pgg2} considered, similarly to
the  behaviour observed in smectic systems, two types of flow:  a horizontal,
outwardly directed radial particle current and vertical permeation fluxes, one
from the neighbouring upper layer and one towards the adjacent layer below.
These fluxes are shown to be important only within a thin annulus of small
size $\xi$ (comparable to $a$) near each step, resembling a ``permeation ribbon''.
Between steps, the
viscous effects associated with simple shear dominate. This leads to simple laws
for the
dilation (or contraction) of the various layers. According to Ref.~\cite{pgg2},
whenever the distinct layers grow laterally at a comparable rate, they all
grow proportional to $\sqrt{t}$ with the proportionality factor
$\sqrt{(W_{n+1} - W_{n})/\zeta_n}$, where $\zeta_n$ is the friction coefficient
between the $n$th and the $(n-1)$th layer. This sheds light on the conditions at
which ``terraced wetting'' can occur. Such a phenomenon takes place if in
several layers close to the substrate the ratios $(W_{n+1} - W_{n})/\zeta_n$
have approximately the same value. If, however, the film closest ($n = 1$) to
the solid substrate grows much faster than all the other layers above, i.e.,
if it decouples from the rest of the drop which then acts as a reservoir for
the first layer, this model predicts that the latter one grows proportional to
$(t/ln(t))^{1/2}$ \cite{pgg2}. This is somewhat slower than $\sqrt{t}$, and
terraced wetting does not occur in such a situation.

Within the classical framework of non-equilibrium statistical mechanics an
alternative description has been worked out by Abraham \textit{et al}
\cite{joela,joelb} and De Coninck \textit{et al} \cite{joelc}. Within this
approach, an interfacial model for the non-volatile fluid edge has been
developed and analysed in terms of Langevin dynamics for the displacements of
horizontal solid-on-solid (HSOS) layers $\{l_{j}\}$ at increasing
heights $j = 0,1, ... $ from the substrate [see
Fig. \ref{layered_models}(b), where $l_j$ is denoted as $h_j$]. These have
essentially the same meaning as the layers appearing in the de Gennes-Cazabat
model \cite{pgg2}, and $l_{j}$ can be thought of as the radius of the $j$th
layer. The radii and the heights are measured in units of $a$ and thus they
are dimensionless.

In the model developed in Refs.\cite{joela,joelb,joelc} one has considered
the case
$f(j) = \delta_{j,0}$ [see Fig.~\ref{layered_models}(b)], where $\delta_{j,0}$
is the Kronecker-delta, so that the energy $U(\{l_{j}\})$ of a given
configuration $\{l_{j}\}$ of a one-dimensional interface is described by
\begin{equation}
\label{U_RHSOS}
U(l_{0}, l_{1}, ... , l_{L}) \; = \; \sum_{j = 1}^{L} P(l_{j} - l_{j-1}) \; -
\; \mu_{0} \; l_{0},
\end{equation}
where $l_{0}$ is the linear extent (in units of $a$) of the ``precursor''
film next to the
solid substrate, $\mu_{0}$ is a wall contact potential and the function
$P(l_{j} - l_{j-1})$ describes the free energy contribution due to the
non-planar liquid-vapour interface. Explicitly, this latter contribution
corresponds to a discretized version of the length of the one-dimensional
interface:
\begin{equation}
\label{en}
P(l_{j} - l_{j-1}) \; = \; J \; \sqrt{1 \; + \; (l_{j} - l_{j-1})^{2}},
\end{equation}
where the parameter $J$ is the surface tension of the one-dimensional
interface.

The dynamics of the layers $\{l_{j}\}$ has been described
in Refs.\cite{joela,joelb,joelc} by a set of $L$ coupled Langevin
equations (recall that $l_L$ is fixed),
\begin{equation}
\zeta \; \frac{\partial l_{k}}{\partial t} \; = \; - \;
\frac{\partial U(\{l_{j}\})}{\partial l_{k}} \; + \; f(l_{k};t),
\end{equation}
where $\zeta$ is a certain phenomenological ``friction'' coefficient per unit
length of the interface, which is supposed to be the same for all layers, and
$f(l_{k};t)$
is Gaussian white noise. Note that in this model $\zeta$ has a different meaning
as compared
to the model by de Gennes and Cazabat \cite{pgg2},
where $\zeta$ is the friction coefficient
between adjacent layers and varies with the distance from the substrate; here
$\zeta$ is associated with the dynamics of the interface only, it does not
depend on the distance from the substrate and it does not account for the
dissipation processes occurring in the bulk.

Note as well that the Langevin description of the interface dynamics tacitly
presumes that $\{l_j\}$ are continuous variables so that the difference
$l_{j} - l_{j-1}$ is not necessarily constrained to be an integer. This implies
that one may encounter two different types of behaviour depending on
whether the difference $l_{j} - l_{j-1}$ is small or large, because the free
energy contribution due to the non-planar
liquid-vapour interface in Eq. (\ref{en}) exhibit different asymptotic
behaviours. Indeed,
in the limit $l_{j} - l_{j-1}  \ll 1$ the function in
Eq. (\ref{en}) is quadratic, i.e., $P(l_{j} - l_{j-1}) \; \approx \; J \;
\{1 + (l_{j} - l_{j-1})^{2}/2\}$, while for  $|l_{j} - l_{j-1}|  \gg 1$ one
has $P(l_{j} - l_{j-1}) \; \approx \; J \; |l_{j} - l_{j-1}|$.

This model allows for an analytical analysis, which
does yield the extraction of a precursor film as well as ``terraced'' forms of
the dynamical thickness profiles. It predicts that for $\mu_{0} > J$ and for
sufficiently short precursors (such that the dominant contribution to the
surface energy is quadratic), the length of the film increases with time as
\begin{equation}
l_{0}(t) \; \sim \; \sqrt{\frac{(\mu_{0} - J) \; t}{\zeta}},
\end{equation}
which resembles the experimentally observed behaviour. For sufficiently
long precursors, for which the surface energy increases linearly with the
length, the  layer next to the substrate grows faster:
\begin{equation}
l_{0}(t) \; \sim \; \frac{(\mu_{0} - J) \; t}{\zeta}\,.
\end{equation}
If $\mu_{0}  =  J$, the precursor film advances as
\begin{equation}
l_{0}(t) \; \sim \; \sqrt{t \; \ln(t)}.
\end{equation}
Therefore, this model predicts that at very large times the advancing
precursor film attains a constant velocity so that the $\sqrt{t}$-behaviour
is only a transient. This inconsistency with the experimental findings can
be traced back to the fact that, by focusing on the evolution of the
interface and assuming a viscous-type dissipation only at the locus of the
interface, the above model neglects the energy dissipation for each molecule
within the precursor film as well as dissipation appearing due to viscous flow
in the spreading droplet.

\begin{figure}[!htb]
\centering
\includegraphics[width= .7 \linewidth]{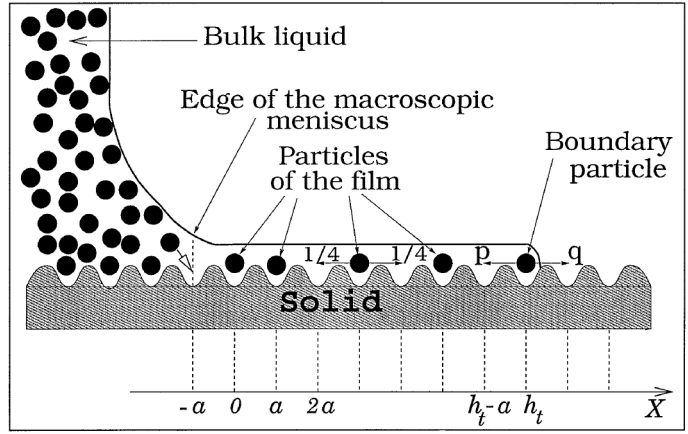}
\caption{
\label{fig_lattice_gas}
Schematic drawing of a molecularly thin precursor film spreading in capillary
rise geometries, i.e., a vertical two-dimensional  wall immersed in a liquid bath as considered in Refs. \cite{bura,burc,burd,osh1,osh2}. In this effectively one-dimensional geometry the $X$-coordinate measures the height above the edge of
the macroscopic meniscus of the liquid-gas interface and $h_t$ defines the
position of the rightmost particle of the film, i.e., of the boundary particle. The
latter is subject to a retentive force due to the presence of a HSOS interface
bounding the film so that the hopping probabilities $p$ and $q$ of the
rightmost particle are asymmetric ($p \leq q$) and are defined by
$p/q = \exp(- \beta J)$. Within this model the particles in the film are not subject
to any mean force and have equal hopping probabilities for jumping to and
away from the meniscus (for a square lattice these probabilities are taken equal
to $1/4$).
[Fig. 1 in {\it Microscopic Model of Upward Creep
of an Ultrathin Wetting Film}, S. F. Burlatsky, G. Oshanin, A. M. Cazabat,
and M. Moreau, Phys. Rev. Lett. \textbf{76}, 86 - 89
(1996) (doi: 10.1103/PhysRevLett.76.86). Copyright \copyright 1996, reprinted
with permission from the American Physical Society.]
}
\end{figure}

To resolve this inconsistency, Burlatsky \textit{et al} \cite{bura,burc,burd} proposed a microscopic,
stochastic model for the spreading of molecularly thin precursors films.
Within their approach the film was considered as a two-dimensional hard-sphere
fluid with particle-vacancy exchange dynamics [see Fig. \ref{fig_lattice_gas}].
Attractive interactions among the particles in the precursor film were not
included into the model explicitly, but introduced in a mean-field-like way.
It was assumed that the film is bounded by a HSOS-type interface
\cite{joela,joelb,joelc}, in which the energy parameter $J$ penalizing non-flat
interface configurations was considered as a certain (not yet specified in
Refs.\cite{bura,burc,burd}) function of the amplitude of the particle-particle
attractions.
\footnote[2]{
\label{foot11}
In Refs. \cite{osh1,osh2} an attempt has been made to express the parameter $J$
in terms of the parameters of the interaction between the fluid particles.
}

The film was taken to be connected to an unlimited reservoir mimicking the
bulk liquid (or a macroscopic drop). The rate at which the reservoir releases
particles into the film has been related to the local particle density in the
film near the nominal contact line and to the strength of the attractive
van der Waals
interactions between the fluid particles and the substrate, in
accordance with the standard Langmuir adsorption theory. In contrast to
Ref. \cite{pgg2}, the model in Refs. \cite{bura,burc,burd} emphasizes
compressibility and molecular diffusion at the expense of hydrodynamic flows.
The latter model assumes that the reservoir and the film are in mechanical
equilibrium with each other, so that there is no hydrodynamic pressure difference
to drive a flow of particles from the reservoir and to push the particles along
the substrate and thus away from the droplet. This approach predicted the
$\sqrt{t}$-law [Eq. (\ref{sqrt})] for the late time stages of the growth of the
molecularly thin film. This law was obtained for the capillary rise geometry and it
was found that the density in the film varies strongly with the distance from
the reservoir.

Moreover, in Refs. \cite{bura,burc,burd} the conditions were established
under which spreading of the precursor film takes place. It was also
suggested that the physical mechanism underlying the $\sqrt{t}$-law is
provided by the diffusive-type transport of vacancies from the edge of the
advancing film to the contact line, where they perturb the equilibrium between
the macroscopic drop and the film and get filled with fluid particles from the
drop.

An extension of the lattice gas model developed in
Refs. \cite{bura,burc,burd,osh1,osh2} has been proposed and analyzed in
Ref. \cite{Popescu2004} which includes an attractive contribution to the
pair interaction between fluid particles, resembling a Lennard-Jones (LJ)
potential. The spreading of the monolayer from a reservoir covering a half
plane has been studied as a function of the strength of the attractive
pair interaction $U_0 =: W_0 k_B T$ among the particles. The study
used both
continuous time kinetic Monte Carlo (KMC) simulations as well as a non-linear
diffusion equation for the local coverage  $c = c(\mathbf{r},t)$
of the monolayer which is the
probability to find at time $t$ a fluid particle in a small area centered at
$\mathbf{r}$:
\begin{equation}
 \label{non_linear_diff}
\frac{\partial c}{\partial t} = \nabla[D(c)\nabla c],~~
D(c) = D_0 [1- G W_0 c (1-c)] \,.
\end{equation}
This equation has been derived, within a mean-field approximation, from the microscopic dynamics
in the corresponding continuum limit. $G$ is a numerical factor which depends
on the lattice structure and on the detailed form of the pair interaction,
while $D_0 = D(c = 0)$ denotes the one-particle diffusion coefficient on the
bare substrate.
We note that Eq. (\ref{non_linear_diff}) has the same mathematical structure as
Eq. (\ref{spr}), which describes spreading of mesoscopically thin films so that the former one can
be \textit{formally} thought of as being
the microscopic analogue of the latter one. However, they
differ in their physical origin. In Eq. (\ref{non_linear_diff}) the coverage
dependence of diffusivity arises from the fluid-fluid interactions and the functional form of $D(c)$ encodes the competition between the normal diffusion due to
concentration gradients and a local ``drift'' induced by the mean-field potential
due to the particle-particle interactions. In contrast, the diffusivity in Eq.
(\ref{spr}) reflects the driving force of the gradients in disjoining pressure,
which encodes the effective interaction between the liquid-vapour interface and the substrate.

\begin{figure}[!htb]
\centering
\includegraphics[width=1.\linewidth]{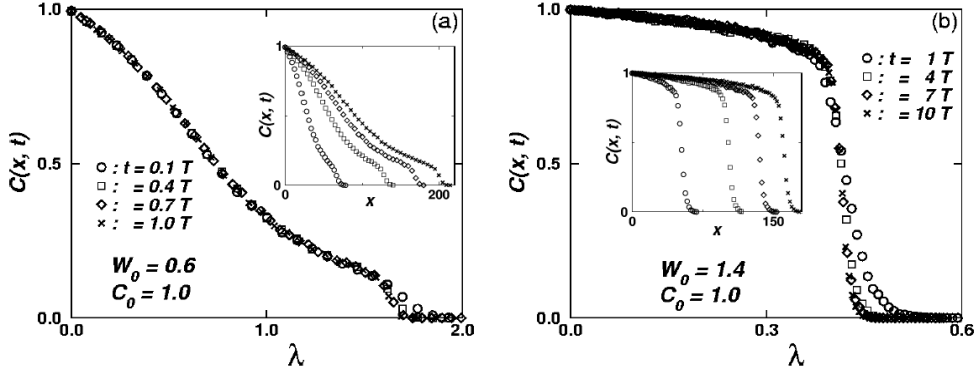}
\caption{
\label{fig_Mihailpaper}
Coverage profiles $c(x,t)$ as functions of the scaling variable
$\lambda = x/\sqrt{D_1 t}$ for values of $W_0$ {\bf (a)} below and {\bf (b)} above
the threshold value for the emergence of sharp interfaces within the film.
The insets show these profiles as functions of the distance $x$
from the edge of the particle reservoir (measured in
units of the lattice constant). The time scale $T = 2 \times 10^6$ is
measured in
units of the average time for a jump attempt of an isolated fluid particle.
[Figs. 6(a,b) in {\it Model for spreading of liquid monolayers},
M.N. Popescu  and S. Dietrich, Phys. Rev. E \textbf{69}, 061602 (2004)
(doi: 10.1103/PhysRevE.69.061602). Copyright \copyright 2004, reprinted
with  permission from the American Physical Society.]
}
\end{figure}

The model confirms the time dependence $\ell_t = \sqrt{D_1(W_0, c_0)\, t}$
of
the spreading with a prefactor depending on the strength of the interparticle
attraction and on the fluid coverage $c_0$ at the reservoir. The
behaviour of
$D_1(W_0, c_0)$ agrees with the existence of a covering-noncovering
transition separatrix $W_0^{(tr)}(c_0)$ in the $(c_0, W_0)$ plane
\cite{osh1,osh2} below which a macroscopic film is extracted from the reservoir
and spreads over the substrate, while above it is not extracted. This
refers only to the formation of a liquid-like monolayer but does not
exclude the formation of a very dilute two-dimensional gas covering
the substrate. The coverage of the expanding film as a function of time
and distance from the edge of the particle reservoir exhibits a scaling
behaviour as function of the scaling variable $\lambda = x/\sqrt{D_0 t}$.
These scaled density profiles exhibit qualitatively different structures above
and below a threshold value of the interparticle attraction; in particular,
above the threshold interaction sharp interfaces form inside the extracted
monolayer (see Fig. \ref{fig_Mihailpaper}).
Such coverage profiles with sharp interfaces are very similar to the
experimentally observed ones for the spreading of Bi on Cu(111)
[Sec. \ref{metal_metal_exper}], but the cause for their formation is seemingly
different. In the former case, they occur because the tendency for clustering
becomes dominant for strong inter-particle attraction, which manifests itself
as an instability (i.e., a negative diffusion coefficient for a certain range
of density values) in the non-linear diffusion equation describing the
spreading, while in the latter case a similar instability occurs due to
surface alloying of the spreading film.

An interesting lattice gas model for the spreading of ultrathin precursor
films, which can be considered as a microscopic version of the continuum
model of permeation layers of de Gennes and Cazabat \cite{pgg2}, is the one
proposed in Refs. \cite{Abraham1994,Abraham2002}.
This is an Ising model with nearest neighbour interactions in an external
field, provided by the substrate potential, on a cubic lattice of infinite
extent along the $x$ and $y$ directions and finite extent along the positive
$z$-direction (in Ref. \cite{Abraham2002}, in units of the lattice spacing
$z \in \{1,2\}$), with local coverage $c(x,y,z) = 0, 1$ if the site is empty
(``vacancy'') or occupied by a ``particle'', respectively. (The sites
$z \leq 0$ are attributed to a continuum substrate and cannot be occupied
by the fluid.) The model is defined by a configuration-dependent energy
at time $t$ given by ${\cal H} = - U_0
\sum_{|\mathbf{r}-\mathbf{s}| = 1} c(\mathbf{r},t) c(\mathbf{s},t)
+ A \sum_{z = 1,2} c(\mathbf{r},t)/z^3$, $A < 0$. The first term corresponds
to a strong nearest-neighbour attraction ($U_0/(k_B T) \gg 1$ in order to ensure
low volatility) and the second one corresponds to a van der Waals attraction by the
substrate. The dynamics is defined through particle-vacancy exchange rates
corresponding to a Kawasaki dynamics augmented by an instantaneous
vacancy-particle-exchange process for a particle which would come to be
located in the upper layer $z = 2$ exactly above a vacancy in the lower layer
$z = 1$. The simulation of the spreading of a precursor, e.g., along the
$y$-direction, employs as initial state a lattice which is half-occupied by
fluid, $c(x,y \leq 0,z, t = 0) = 1$, with a fixed coverage $c(x,y = 0,z,t) = 1$
mimicking a fast relaxation of the liquid reservoir at the edge connected
to the film. The particles move away from the line $y=0$ into the initially
empty area solely due to the concentration gradients; there is no forcing
by the macroscopic film occupying the half space $y < 0$. The advancing
edge $\ell_t$ is defined in the $z = 1$ plane as the largest $y$ position among the rightmost
particles in the maximal connected the cluster, i.e., the
percolating one in contact with the reservoir.
Based on kinetic Monte Carlo simulations it was concluded that the
model predicts a compact, first layer precursor spreading which follows
$\ell_t \sim \sqrt{t}$ and the emergence of a second layer with a much
slower spreading. The mechanism of spreading is the diffusion in the lower
layer of a low concentration of vacancies, from the advancing edge towards
the ``reservoir''. This is coupled to the diffusion in the upper
layer of a low concentration of particles in the opposite direction,
as well
as to the vacancy-particle exchange between the two layers. Therefore,
this model points towards a promising direction for studying terraced spreading, which requires
simply to relax the constraint $z \leq 2$. However, it is difficult to estimate
the ensuing impact on the computational cost (see also Ref. \cite{Abraham1994}).
Moreover, it is unclear whether a non-uniform structure of the first layer,
which would involve particle concentration variations along the spreading
direction (see, e.g., Ref. \cite{Moon2001}) is compatible with the rules
defining the model. This issue arises because the diffusion of vacancies
through the high coverage lower layer is much slower than that of particles
in the very low coverage upper layer. Thus with increasing extent  $\ell_t$ of
the precursor (i.e., the first layer) it is likely that the vacancy-particle
exchange mechanism becomes very effective, leading to a vanishing density of
vacancies in the first layer.

\section{\label{sec_num} Numerical studies of ultrathin precursors}

Computer simulations of various models for liquid droplets spreading on
solid substrates have been a very valuable tool in the quest for understanding
the mechanisms behind the emergence of terraced spreading and the dynamics
of ultrathin precursor films. The methods, which are usually employed,
encompass Monte Carlo (MC), kinetic Monte Carlo (KMC), and Molecular Dynamics
(MD). They are discussed in detail in many textbooks (see, e.g.,
Refs. \cite{Binder_book,Frenkel_book}). Therefore here we focus strictly on
their use for studies of precursor film spreading.

\subsection{\label{sec_MC} Monte Carlo simulations}
Monte Carlo simulations have been employed to elucidate the emergence of a
precursor film (first layer) with a growth behaviour $\ell_t \sim t$ (which
disagrees with the experimental observations) within the so-called horizontal
solid-on-solid (HSOS) model of non-volatile liquid droplet spreading
(see Sec. \ref{sec_models}). As discussed above, this model can be
solved exactly in certain limiting cases. But although it captures qualitative
features of the available experimental results, such as terraced spreading,
it predicts a too fast dynamics for the first layer acting as the precursor
film. The question arises whether this situation is generic for all the values
of the surface tension $J$ and of the parameters characterizing the interaction
with the substrate [see Eq. (\ref{U_RHSOS})]. If true, this would rule out
this model. Otherwise, the too fast dynamics could be a consequence of the
approximations made in order to obtain analytic solutions. Monte Carlo
simulations of the HSOS model for a liquid wedge have been performed both in
two and in three dimensions (i.e., considering a straight edge or allowing for
a fluctuating edge for each molecular layer, respectively) for various values
of the surface tension and for various cut-off ranges of a van der Waals type
substrate potential which decays as $1/z^3$ with the distance $z$ above the
planar continuum substrate \cite{Abraham1991_jpa,Abraham1991_pa,Abraham1992}.
Except for special values of the parameters (e.g., above, but very near, to
the wetting transition line, i.e., close to the temperature at which complete
wetting occurs) for which the results are not conclusive due to the very
slow dynamics, the simulations provided strong evidence that the first layer
indeed spreads with a linear time dependence. This demonstrates that the model
is indeed too simplistic. Nonetheless, the time dependence should be
interpreted with some caution because these MC simulations have followed the
dynamics in ``computer time''. This means that each trial move is assumed to
take the same average time in order for the time scale of MC steps per lattice
site to acquire a physical meaning. This differs from the real time provided by
the use of transition rates for various processes, as required by a rigorous
KMC method \cite{Binder_book}. Later it has been argued that the most likely
cause for this discrepancy with the experimental observation is that the model
does not account for the energy dissipation associated with the necessary
transport of mass from the edge of the drop to the advancing tip of the
precursor \cite{Abraham2002}.

A three-dimensional Ising model for a droplet spreading upon
contact with a planar substrate was proposed by Abraham \textit{et al}
\cite{Abraham1995}. In this cubic lattice model, the flat substrate
occupies the half space $z \leq 0$, while the sites with $z > 0$
can be either occupied by a fluid particle ($c_i = 1$) or a void
($c_i = 0$), where $c_i \in \{0,1\}$ denotes the occupation of site $i$.
The model employed a fixed size of the simulation cell (therefore there is
a fixed volume available to the fluid) and a fixed number of fluid particles. (These
constraints do not play a major role in the case in which three-dimensional
evaporation is negligible, as it was the case for the simulations in
Ref.\cite{Abraham1995}, see below.) The initial state is that of a
lattice approximation of a hemi-cylindrical drop spanning the whole
simulation box along the
$y$-direction, for which periodic boundary conditions have been
imposed. The spreading takes place along the $x$-direction, and the spreading
is studied at times for which the advancing edge of the drop or of the
precursor (see below) remains far from the boundaries of the simulation box
along the $x$-direction. There is an attractive interaction of
strength $W_0$ between nearest-neighbour fluid particles, and the substrate
interacts with the fluid via a van der Waals type potential. Therefore the
corresponding Hamiltonian
\begin{equation}
\label{Ising_Abraham}
{\cal H}
= - \frac{U_0}{2} \sum_{\langle i,j \rangle} c_i c_j  +
A \sum_{i} \frac{c_i}{z_i^3},~~U_0 > 0\,, A < 0,
\end{equation}
where $\langle i,j \rangle$ denotes nearest-neighbour sites $i$
and $j$ and $z_i > 0$ is the distance of site $i$ from the substrate at
$z = 0$ in units of the lattice spacing.
The MC simulations of this model employed a dynamics specified to be a
particle-void exchange with Kawasaki rates (see, e.g.,
Ref. \cite{Binder_book}). By varying the ratio $U_0/A$ of the strength of
the particle-particle and of the particle-substrate interactions, while
keeping the thermal energy sufficiently low in comparison with the
particle-particle interactions strength ($k_B T/U_0 = 1/W_0 \leq 2$),
such that three-dimensional evaporation was negligible, it was shown that
the model
exhibits the emergence of terraced spreading. The number of steps and their
rate of spreading turned out to be practically independent of the strength
of the substrate potential. After a transient time, which can be rather
long if the substrate potential is not strong, the spreading of each layer
obeyed the $\sqrt{t}$ dependence on the ``time'' measured in MC
steps per lattice site. In this case, the mechanism of mass transport to the
film was shown to be a flux of particles along the terraced surface of the
drop.  This is the microscopic equivalent of the permeation layer in the
continuum model proposed in Ref. \cite{pgg2}. However, the maximum lateral
extents of the precursor films observed in these simulations are still
relatively small (no more than $60$ to $80$ lattice sites, according to
Figs. 1 and 2 in Ref. \cite{Abraham1995}, compared with an initial
footprint of the
drop of about $30$ to $40$ sites and a height of about $10$ sites).
Therefore it
is difficult to assess if they can indeed be associated with the
\textit{macroscopically} large molecular films observed experimentally.
By varying the ratio $k_B T/U_0$ the role of evaporation could be studied.
The conclusion was that the qualitative features of the formation and
dynamics of the precursor remain the same even if significant
evaporation occurs. By varying the number of particles in the drop between $500$ and
$24000$, also the effects of the drop size on the dynamics have been
investigated. It was found that there is no significant difference in the
behaviour as long as the drop is far from being emptied by the spreading
precursors.

A coarse grained description, which emphasizes the mass transfer along the
surface of the drop, has been proposed by De Coninck \textit{et al}
\cite{DeConinck1993_pre,DeConinck1993_la}. This is a restricted
solid-on-solid (RSOS) model in which the drop and the film (phase A) in
contact with the flat substrate (W) and a gas phase (B) are described by
a  set of columns with height values $\{l_0,l_1,\dots\}$ at discrete
positions $\{x_i,y_i\}$ on a planar regular lattice, assuming that there
are no overhangs. The free energy of a configuration comprises
contributions proportional to the areas (or, in a two-dimensional version,
the perimeter) of the liquid-vapour, liquid-wall and vapour-wall contact
multiplied with the corresponding surface tensions, as well as an
additional term $\mu(l)$ modeling a long-ranged effective interaction of
the liquid-vapor interface with the substrate; $d\mu/d l$ is the
so-called disjoining pressure. The dynamics is introduced as transfers
of at most one unit height (thus the name ``restricted'' SOS model) between
neighbouring columns with Kawasaki type rates. Due to the coarse grained
nature of the model the MC simulations span much larger time- and
length-scales than the ones for the microscopic Ising model discussed
above. The results show that the model captures the emergence of terraced
spreading \cite{DeConinck1993_la} as well as spreading of a single layer
ahead of the drop \cite{DeConinck1993_pre}. The thickness of these
layers depends on the range and the strength of the effective
interface potential $\mu(l)$. In all cases, the dependence on time (measured
in units of MC steps per site) of the spreading of
the precursor is reported to be in agreement with the expected
$\sqrt{t}$-law.

As discussed in Sec. \ref{sec_models}, in
Refs. \cite{Lacasta2001,Popescu2004} a lattice gas model for the spreading
of a monolayer precursor in contact with a reservoir of particles mimicking
a drop \cite{burc,osh2,Popescu2004} was studied by using continuum time KMC.
Due to the focus on the dynamics of the precursor, the computer
simulations could explore the dynamics on relatively long time (seconds)
and large spatial scales of more than a hundred times the molecular size.
They confirmed that the model predicts that the linear extent of the
spreading monolayer
(provided it does occur) grows in time as $\sqrt{t}$. These simulations
revealed a complex structure of the concentration profiles as a function
of the distance from the reservoir, including the emergence of sharp
interfaces within the expanding monolayer. A similar
model, but one in which the constraint of single-occupancy of a site was relaxed,
has been proposed for studying the spreading of a mesoscopically thick
precursor films \cite{Dotti2007}. Numerical simulations of this model in a
one-dimensional geometry predict a spreading dynamics of the emerging precursor
film which also follows a $\sqrt{t}$ dependence on time.

Finally, we note that in Ref. \cite{Milchev2002} MC simulations have been
reported for the spreading of a polymer droplet of chain-like molecules.
They consist of $32$ particles with a finitely-extendible, nonlinear and
elastic
potential between the components of each chain, with a Morse potential as
the
pair interaction between the components of distinct chains, and with an
attractive van der Waals particle-substrate potential. While macroscopic
drop spreading (i.e., Tanner's law) and the early stages of precursor
formation have been successfully captured by the simulations, the spreading
of the precursor film could not be studied in detail because due to
computational limitations the drop size of $128$ chains with $32$ particles
each was too small. Such numerical limitations are expected to become
severe if the number of conformational degrees of freedom per molecule is
large, in particular as the complexity and the range of the pair potentials
are increased.

\subsection{\label{sec_MD_drop} Molecular Dynamics simulations for simple
or polymeric liquids}
A molecular description is provided by MD simulations, in which the
particle-particle interactions are specified and the dynamics
follows from direct integration of Newton's equation of motion. In the
context of droplet spreading and the formation of precursor films, MD
simulations have been a very useful tool in obtaining direct insight into
the molecular details of the physical mechanisms driving the spreading (see
Refs. \cite{DeConinck1996_cs,Voue2000,Grest2003,Samsonov2011} and references
therein). Before embarking on a summary of the main corresponding developments,
a word of caution is yet in order. While the MD method has the advantage of
eliminating from the modeling part the numerous assumptions needed as input
for the rates which define a MC or KMC dynamics, the method suffers from being
computationally extensive and therefore requiring extremely large computer
resources (both in terms of memory and CPU time) even for simulating relatively
small systems (100 000 particles) over real-time scales of less than one
nanosecond. In addition, the actual interaction potentials are rarely well
known. The results of MD simulations of droplet spreading should thus be
considered as capturing qualitative features, rather than providing a
\textit{bona fide} quantitative description. Moreover, they should be
carefully scrutinized with respect to finite size effects and time scales
which are intrinsically small, in particular if an interpretation for the
asymptotic long time behaviour is sought (for an illuminating discussion see,
e.g., Refs. \cite{Frenkel_book,Grest2003}).

The first MD studies successfully captured the occurrence of terraced
spreading but led to contradictory results for the dynamics of precursors
\cite{Koplik1991,Koplik1992,Nieminen1992,Nieminen1994}. In
Refs. \cite{Koplik1991,Koplik1992} fully atomistic MD simulations have been
performed, in that both the drop and the substrate were represented as
atoms with corresponding inter-particle interactions of a Lennard-Jones
form with a cut-off range of the order of the atomic diameter. The various
potential parameters characterize the fluid-fluid, fluid-solid, and
solid-solid pair interactions, and they are chosen such that at the
temperature of the simulations most of the fluid is in the liquid state and
that the solid maintains its initial crystalline fcc structure with a (100)
plane exposed to the liquid. Due to the aforementioned computational
constraints, the size of the system was very small: 4000 particles for the
fluid and 9000 particles for the solid, which corresponds to five atomic
planes. By varying the strength of the liquid-solid interactions, while
keeping the fluid-fluid ones fixed, the simulation explored various wetting
regimes. In Ref. \cite{Koplik1992} an additional strong bonding was introduced
between pairs of fluid particles, mimicking diatomic molecules; this sheds
light on the influence of the size of the fluid particles on terraced
spreading. The results provide clear evidence for the occurrence of terraced
spreading as well as for layering inside the core of the drop but still
within the liquid state. As intuitively expected, the number of distinct
layers
increases with increasing strength of the solid-fluid interactions. In these
MD simulations the atomic liquid was fairly volatile, whereas the diatomic
molecules were practically non-volatile. The simulations revealed that for
very strong substrate-fluid interactions the vapor played some role in the
formation of the precursor, but by comparing with the case of diatomic
molecules it turned out that this evaporation-condensation process is a
sub-dominant effect. However, in all cases studied the precursor film
associated with the liquid layer next to the substrate showed a much slower
spreading, i.e., $\ell_t \sim \sqrt{\ln(t)}$, than the one observed
experimentally and predicted theoretically. This result was puzzling, as it
apparently was not a finite-size effect: simulations with twice as many fluid
particles reproduced the same behaviour \cite{Koplik1992}.

The MD studies in Refs.\cite{Nieminen1992,Nieminen1994} considered the case
of an atomic fluid as well as that of a binary mixture of single particles,
acting as a solvent, and chain molecules. The latter consist of two, four,
or eight single particles interconnected by a stiff, isotropic harmonic
oscillator potential. All particles are taken to interact via Lennard-Jones
potentials and they are in contact with a homogeneous impenetrable substrate
which additionally interacts with the particles at a distance $z$ from the
substrate via a van der Waals type potential $A/z^3$, $A < 0$. In
Ref. \cite{Nieminen1994} the impenetrability condition is implemented by a
strongly repulsive term in the substrate-particle interaction
$A/z^3 + B/z^9,~ A < 0,\,B > 0$. This substrate potential follows from
integrating over a half-space the pair potentials between substrate and fluid
particles. Concerning the leading term this is equivalent to carrying out the
corresponding discrete sum of pair potentials as used in
Refs. \cite{Koplik1991,Koplik1992}. The simulations, performed at temperatures
at which evaporation is negligible, have shown that for atomic or diatomic
molecules a precursor film occurs in most of the cases, while for longer molecules
with orientational degrees of freedom layering and terraced spreading occur
only if the strength of the attractive part of the substrate potential is
above a threshold value which depends on the chain length. In contrast
to the results reported before \cite{Koplik1991,Koplik1992}, in all
cases in which a precursor film occurred its dynamics did show a $\sqrt{t}$
spreading behaviour after a transient due to the precursor formation. 
This was followed by a slower expansion once the reservoir drop begins 
to empty. (In Ref. \cite{Nieminen1992} it is pointed out explicitly that 
the initial spreading with nearly constant speed later crosses over to diffusive 
spreading, in turn followed by a slower regime once the macroscopic drop has 
emptied into the film.) 
Concerning the studies in Refs. \cite{Koplik1991,Koplik1992}, the discrete
structure of the substrate consisting of particles of the same size as
the one of the fluid particles, the strong substrate-fluid pair interactions, and the
moisture of the substrate due to significant evaporation-condensation
processes have been surmised \cite{Nieminen1992} to be the most likely
sources for the significant difference between the results of
Refs. \cite{Koplik1991,Koplik1992} and Refs.\cite{Nieminen1992,Nieminen1994}.

The issue of the dynamics of the precursor spreading has been further
investigated with MD simulations in Refs. \cite{DeConinck1995,DeConinck1996_pre}.
These studies employed the same atomistic representation of the substrate as
in Refs.\cite{Koplik1991,Koplik1992} with a cut-off for all Lennard-Jones
pair potentials at a distance equal to 2.5 times the fluid core size, but
with an additional $A/z^3$ contribution to the resulting substrate potential.
However, these studies used chain molecules consisting of 8 or 16 atoms,
bound together by a confining pair potential. This led to negligible
evaporation and eliminated the similarity in size between the species forming
the solid and the fluid, respectively. The simulations revealed the formation
of a well-defined first layer precursor film, as well as up to three additional
layers spreading much slower than the first one. The dynamics of the first
layer shows a clear $\sqrt{t}$ behaviour, which therefore indicates that the
behaviour $\ell_t \sim \sqrt{\ln(t)}$ reported previously must be
related to the size of the fluid particles used there. It was speculated that due
to the compatibility in size the fluid particles in the previous MD simulations
are trapped for long times in the local minima of the corrugated substrate
potential, which leads to a slow motion along the surface and a correspondingly
slow spreading, whereas the chain molecules are incommensurate
with the distribution of minima in the corrugated substrate potential and
therefore can move easily along the surface
\cite{DeConinck1995,DeConinck1996_pre,DeConinck1996_cs}.

\begin{figure}[!htb]
\centering
\includegraphics[width= .8 \linewidth]{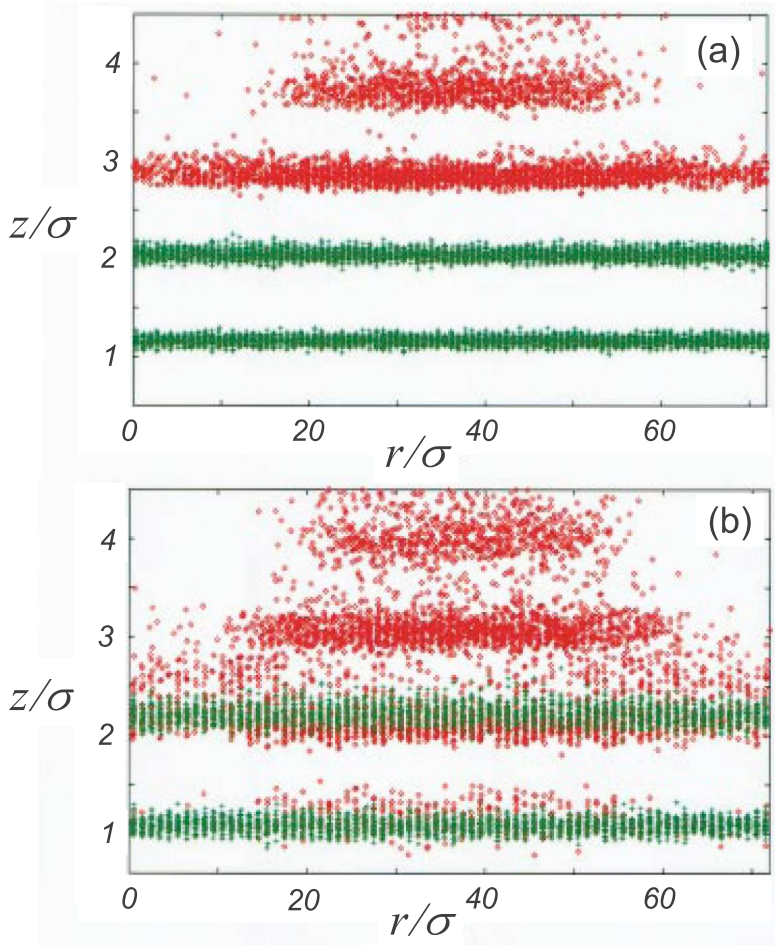}
\caption{
\label{fig_MD_porous}
The two uppermost layers of a solid (green) in contact with a fluid drop
(red) forming layers, too.  In contrast to a solid with nearest-neighbour
distance $d_{NN} =­ 2^{1/6} \sigma$ {\bf (a)} (where $\sigma$ is the
distance at which the fluid-fluid interaction potential vanishes), the
solid with $d_{NN} =­ 2^{2/3} \sigma$ {\bf (b)} allows the fluid particles
to penetrate the solid which thus acts like a porous material.
[Fig. 4(a,b) in {\it Simulating the spreading of a drop in the terraced
wetting regime}, S. Bekink, S. Karaborni, G. Verbist, and K. Esselink,
Phys. Rev. Lett. \textbf{76}, 3766 - 3769 (1996)
(doi: 10.1103/PhysRevLett.76.3766). Copyright \copyright 1996, reprinted
with permission from the American Physical Society.]
}
\end{figure}

The reason for the slow $\sqrt{\ln(t)}$ spreading behaviour has been
finally clearly
identified by the MD study in Ref. \cite{Bekink1996}. It revealed the
substrate  as being porous, which occurs for specific choices of the lattice
constant of the substrate and of the core-size of the fluid particles; they
are both implicitly fixed by the choice of the parameters in the
Lennard-Jones pair potentials of the substrate and the fluid particles,
respectively. The study employed an atomic fluid and a discrete representation
of the substrate as atoms forming a lattice. By varying the parameters in the
various Lennard-Jones pair potentials, the size of the initial drop, the
structure of the substrate lattice [(fcc) or (sc) lattices], and the type
of thermostat employed in the simulations (see Ref. \cite{Frenkel_book}),
the authors convincingly showed that in all situations when the precursor
film occurs (i.e., for a sufficiently attractive substrate potential and
for a sufficiently large drop) its spreading dynamics follows the $\sqrt{t}$
time dependence if the substrate is \textit{impenetrable}
[see Fig. \ref{fig_MD_porous}(a)] and if the drop acts as a reservoir. All
the other parameters (including the evaporation-condensation processes)
influence only the prefactor in this dependence [see Eq. (\ref{sqrt})].
In contrast, for the parameters similar to those used in the studies in
Refs. \cite{Koplik1991,Koplik1992}, the fluid penetrates into the solid
[see Fig. \ref{fig_MD_porous}(b)] and this indeed gives rise to a much
slower spreading which is compatible with a logarithmic time dependence.

Further MD studies focused on the case of chain-like molecules, which is
closer to experimental situations involving the spreading of polymeric oils.
The authors of Ref. \cite{Haataja1995} employed chain-like molecules, one
end of which interacts with the substrate differently than all the other
elements of the chain, as it occurs in the case, e.g., of OH terminated
PDMS. The MD simulations results revealed that
chains which have a preference for being oriented perpendicular to the
substrate (this is controlled by adjusting the interaction
potential between the modified end element and the substrate in such a way
that the minimum of this potential occurs at a distance from the
substrate twice as large as that corresponding to the other elements of the
chain), lead to a more enhanced lateral structure of the precursor film
than chains which prefer to lie flat on the substrate \cite{Haataja1995}.
Although in both cases the dynamics of the
emerging monolayer  precursor films agrees with the $\sqrt{t}$ behaviour,
the prefactor corresponding to the former case is significantly smaller,
which is intuitively expected for an enhanced in-plane structuring.

The role of the particle size and of the substrate potential was
further explored with MD simulations of binary mixtures with an equal number
of chain-like molecules consisting of 8 and 16 atoms, respectively, for which the
interactions with the substrate atoms are chosen such that the 16-atom
molecules are completely wetting the substrate, while the 8-atom
molecules interaction with the substrate is tuned from a partial- to a
complete-wetting behaviour \cite{Voue2000_la}. A precursor film with the
expected $\sqrt{t}$ dynamics is observed in all cases because one component,
i.e., the 16-atom chains, is in the complete wetting regime. The spreading
speed lies between the ones of the pure components. As expected, the
monolayer precursor is enriched by and the next layer depleted of the
16-atom chains as long as the interaction of the 8-atom chains with the
substrate maintains a strength leading to partial wetting.
Similar conclusions have been drawn from recent \cite{Grest2004} much
larger scale MD simulations (300~000 atoms, interacting according
to Lennard-Jones potentials) of hemi-cylindrical droplets consisting of
either pure 10-, 40-, and 100-atoms chains, respectively, or mixtures of
them.

Large scale and long times MD simulations, which allow a simultaneous study
of both droplet spreading and precursor film formation and spreading, have
been performed in order to investigate the role played by various details,
such as the atomic versus a homogeneous structure of the substrate, the type of
thermostat algorithm used, the initial shape of the drop, or the size of
the substrate  \cite{Grest2003}. The conclusion is that the
computationally most effective method is to use a continuum representation
of the substrate, if the porosity of the substrate (see, e.g.,
Ref. \cite{Bekink1996}) or substrate alloying (see, e.g.,
Refs. \cite{Webb2003,Grest2005_prl}) are expected to be irrelevant, and
to use a
thermostat coupling which decays rapidly with the distance from the
substrate. The study in Ref. \cite{Grest2003} reports the formation of
precursor films only for parameters of the LJ pair-potentials which ensure
complete wetting of the substrate by the liquid of chain-like molecules. This
is in agreement with similar observations made in
Refs. \cite{Voue2000,Grest2004}, but it leaves open the question why such
films do not occur in partial wetting situations, which is in contrast to
the experimental observations (see Sec. \ref{metal_metal_exper}) and to the
MD simulations results for metal on metal systems (see, c.f., Sec.
\ref{sec_MD_metal}). Whenever the precursor occurs, its spreading dynamics
\textit{at sufficiently long times} is clearly compatible with a $\sqrt{t}$
time dependence for as long as the drop acts as a reservoir and the advancing
edge of the film is far from the edges of the substrate. The speed of
spreading decreases with the length of the chain-like molecules. Similar
conclusions have been reached in MD simulations of a nano-sized droplet with
LJ pair potentials spreading on a substrate in the complete wetting case
\cite{Sedighi2011}. Additionally, this latter study emphasizes a comparison
with the case of strongly forced wetting occurring when a drop with a
non-zero vertical velocity hits a flat impenetrable substrate. In
this case inertial terms are actually dominant: the drop spreads and
retracts several times before reaching the equilibrium state. Precursor
films do not occur because their kinetics is too slow at the time scales
imposed by the external driving.

The question of the influence of substrate corrugation, i.e.,
the  periodic lateral variation of the substrate potential due to
the crystalline structure of the top layer of the substrate, on the
dynamics of the precursor remains basically open. If one would investigate
the spreading of a colloidal suspension, corrugation could indeed be
completely switched off, because for the big colloids the substrate is
\textit{de facto} not corrugated; however, for MD simulations of
molecular liquids or metal-on-metal systems, this is not the case.
The studies in Refs. \cite{Bekink1996} and \cite{Grest2003} clearly
show two limiting cases, one in which the influence of the substrate
corrugation is very significant (Ref. \cite{Bekink1996}), the other in
which it plays a minor role (Ref. \cite{Grest2003}).  We are not aware
of any systematic MD studies of the dynamics of the precursor as a function
of the lattice structure of the substrate, e.g., concerning the crossover
between porous and non-porous substrates.
The studies in Ref. \cite{Grest2003} indicate that as long as the substrate
is not porous the spreading follows the $\sqrt{t}$.
 However, for the case of non-porous substrates systematic
 MD studies of the effects of the lattice geometry and
 of the lattice spacing on the value of the prefactor are
 not yet carried out.

An interesting approach for resolving the issue of volatility
in MD simulations has been proposed recently in Ref. \cite{Wu2010}, where the
authors use a quarter of a cylindrical liquid drop confined by a
rectangular wedge formed by two atomistic semi-infinite fcc
walls (minus their overlap).  This
quarter of a cylindrical liquid drop is covered by an immiscible fluid
of equal density and viscosity, and the confinement is completed by two
additional homogeneous walls enclosing the covering fluid; periodic boundary conditions apply  in the longitudinal direction. By studying the spreading of the first
liquid  covered by a second, immiscible one, the issue of volatility is
bypassed. Both fluids are modeled as atoms interacting with LJ pair potentials. The
interaction with the atomistic walls is described by LJ
pair-potentials, too, whereas the interaction of any liquid atom with the other
two  homogeneous walls is described by a van der Waals type substrate potential
corresponding to a wedge formed by two semi-infinite walls (minus their overlap). The spreading
of the drop away from the corner of the atomistic wedge is initiated by ``turning on'' an additional attractive van
der Waals substrate potential between the droplet atoms and the bottom,
horizontal and atomistic wall, along which the spreading takes place.
By varying the strength of this latter substrate potential, the
liquid-liquid-substrate system is driven from a partial- to a
complete-wetting state. In agreement with the previous results of MD
simulations for LJ fluids (composed of atoms or chain-like molecules)
\cite{Grest2003,Grest2004,Voue2000,Voue2000_la},
only for the complete wetting case a thin precursor film extending in
front of the ``macroscopic'' edge of the drop  has been observed. The
extension of this film was shown to be in agreement with a $\sqrt{t}$
time dependence, with a prefactor depending on the strength of the
substrate potential. By fitting \textit{ad hoc} a power law to the extent
of this precursor film close to the partial- to complete-wetting transition point -- which is \textit{a priori} unjustified, because in that region
the linear extent of the film over the time scale of the simulation is microscopic in size, i.e., a few atomic
diameters -- it has been claimed that the emergence of precursor films is
an indication of a transition from partial- to complete-wetting.
While this claim seems to hold for LJ fluids and van der Waals substrate
potentials, it cannot be generalized to other fluid-solid models as it
obviously is at odds with the experimental evidence (see
Sec. \ref{metal_metal_exper}). As we shall discuss below this is also at
odds with the results of MD simulations for other pair potentials,
such as for metal-on-metal systems. In these systems ultrathin precursor
films,  expanding proportional to $\sqrt{t}$, emerge also in partial wetting situations. This is intuitively expected in view of the equilibrium picture of
partial wetting corresponding to drops being in spatial contact with microscopically
thin films (see Sec. \ref{sec_intro}).

Maybe the most important recent MD simulation results in this field are
those reported in Ref. \cite{Grest2005_prl}. The dependence of
the precursor spreading on the size of the drop, taken to be a
hemi-cylindrical drop in contact with a homogeneous flat substrate, was
studied in both \textit{complete and partial} wetting regimes. For the
former, a polymer (chain-like molecules with 10 particles) fluid was
used, in contact with a substrate acting on the fluid with an integrated
LJ potential. For the latter, an atomistic representation of Pb on
Cu(111) was used; the equilibrium contact angle was found
to be 33$^\circ$, which is very close to the experimental observations
in Ref. \cite{Moon2001}. (We re-emphasize, though, that an extrapolation
of MD results to macroscopic scales and predictions of material specific properties deserve caveats.) In
both cases the temperature and the parameters defining the potentials
have been chosen such that the fluid is in the liquid state and practically
non-volatile during the time scale of the simulations. The large scale MD
simulations involving up to 78 000 polymer chains and up to 220~000 Pb
atoms have provided the following results:
(i) In both cases a monolayer precursor film emerges out of the
drop and spreads ahead of it. The radius of the precursor film, extending
to more than 100 times the drop radius, follows a $\sqrt{t}$ time dependence
in both cases; for Pb on Cu(111), the total MD simulation time has
corresponded to approximately 4 ns, which is a very long time scale by MD
standards. (ii) In both cases, the prefactor in the $\sqrt{t}$ dependence
is directly proportional to the square root of the initial
radius $R_0$ of the droplet. Extending this finding to the more common case
of a spherical drop, this latter result can be interpreted as a dependence
of the speed of the spreading precursor film on the volume of the drop.
It remains to be seen if such a dependence turns out to be a finite-size
effect due to the intrinsically small spatial scales explored by MD
simulations. This is a new challenge for theoretical analyses, which
so far have been focused on the case of a large \textit{reservoir-like}
drop and a spreading speed which is independent of the drop volume.
It remains as an open question why in these MD simulations
the precursor film occurs even under partial wetting conditions, whereas
in other MD simulations it does not.

\subsection{\label{sec_MD_metal} Molecular Dynamics simulations
for metal-on-metal systems}

The experimental results which have provided evidence for precursor films
in metal-on-metal systems (see Sec. \ref{metal_metal_exper}) have stimulated
the interest in MD simulations for such systems.

Moon \textit{et al} \cite{Moon2002_cms} studied the formation and spreading
of precursor films of Ag on Ni(100). The choice of Ag and Ni was motivated
by the fact that the inter-atomic interactions in metallic systems consisting
of different atomic species can be calculated by using the semi-empirical
embedded atom method (EAM), and that such EAM potentials are well established
for Ni-Ni, Ag-Ag, and Ag-Ni pairs. The MD simulation employed a hemi-cylindrical
Ag drop on an atomistic representation (10 atomic planes) of the Ni(100)
substrate. The formation and spreading of precursor films from the already
equilibrated Ag drop was studied at the temperatures T = 800 K, 900 K and 1000 K.
(At these temperatures the simulated Ag drop is actually in a solid state,
according to the phase diagrams obtained from numerical simulations
\cite{Webb2005}).
At all temperatures, a film with submonolayer coverage emerged. The
coverage $c$ varied as a function of the position $x$ from the edge of
the drop and of the time $t$. The center of the drop maintained a coverage in
between 3 and 4 monolayers, fulfilling its role as a reservoir of particles.
The dynamics of the film is compatible with the $\sqrt{t}$ time dependence,
as shown by the scaling of the coverage profiles when plotted as function of
the scaling variable $x/\sqrt{D_1 t}$. No surface alloying or surface ordering in
the precursor film were observed. The scaled profiles showed complex shapes,
indicating coverage-dependent diffusion coefficients. A Boltzmann-Matano
analysis \cite{Moon2001,Wynblat2007} was used to extract the diffusion
coefficient as a function of coverage. The results exhibit qualitative
features of the diffusion coefficient such as a minimum at intermediate
coverages and a maximum at large coverages. This behaviour shares a certain
similarity with corresponding predictions of lattice-gas models with particles
interacting via simple LJ potentials as proposed in
Refs. \cite{bura,burc,Popescu2004}.

The study of Ag spreading on Ni substrates has been extended by Webb
\textit{et al} \cite{Webb2009} to include the case of cylindrical liquid Ag
droplets brought in contact to and spreading on Ni(111) or Ni(100) surfaces.
The MD simulations have been performed at 1200 K and 1500 K, i.e., slightly
above and well above the melting point, respectively. The experimental bulk
melting temperature of Ag is 1235 K, whereas the MD predictions based on
the EAM potentials used in Ref. \cite{Webb2009} is 1144 K. The numerical
results show that spreading on the two surfaces is practically independent
of the (111) or (100) surface structure. At 1200 K as a final state incomplete
wetting with a contact angle of ca. $10^\circ$ has been found. At 1500 K a
significant dissolution of the Ni substrate occurred and no precursor film
has been observed. This is similar to the results of MD simulations for
high temperature Ag drops on a Cu surface \cite{Webb2005}). In the
simulations at 1200 K no surface alloying and very little substrate dissolution
were observed, together with evidence for the formation of a submonolayer
precursor film. However, even at -- by MD simulations standards long -- times
of approximately 12 ns the extent of the precursor is extremely small. By
analyzing the data only up to 4 - 6 ns one could easily be misled to conclude
that no precursor is formed. This further emphasizes the need for extreme care
in extrapolating the results of MD simulations towards macroscopic, asymptotic
behaviours.

A MD simulation study of precursor films of Pb emanating from liquid Pb
droplets in contact with Cu(111) and Cu(100) substrates at 700 K has been
reported in Refs. \cite{Webb2003,Grest2005_prl}. The precursors occur in a
state of partial wetting (contact angles of 33$^\circ$ and 18$^\circ$ on
Cu(111) and Cu(100), respectively).  In both cases the precursor films expand
by following a $\sqrt{t}$ time dependence, with a significantly larger prefactor
for the Cu(111) substrate. On Cu(100) the precursor is a submonolayer
thick and, in agreement with theoretical expectations \cite{Prevot2000},
significant surface alloying is observed. In contrast, on Cu(111) no
alloying is observed and the precursor consists of two layers. While the lack
of surface alloying seems to contradict the theoretical expectations and the
experimental observations \cite{Moon2001,Humfeld2004}, the occurrence of
films thicker than a monolayer seems to agree with the experimental observation
of a thick film in equilibrium with drops at temperatures above the melting
point of Pb \cite{Moon_la2004}.

The results and studies discussed in Section \ref{sec_MD_drop} have
significantly contributed towards the understanding of the microscopic
mechanisms of ultrathin precursor film formation and of the structure of
these films. They
have helped to reveal effects particularly significant for small drops, such
as the slowing down of a precursor when the size of the drop becomes too
small to act as a reservoir. This also holds for the apparent absence of
precursor film formation over time scales which are long by MD standards,
yet too short to be asymptotically long. This demonstrates that great care
must be exercised concerning the extrapolation of MD data to the large
spatial scales and the long time scales typically explored by experimental
studies. These observations also tell that, after all, it is only the
careful comparison with experimental data which validates a numerical
simulation. Moreover, even in the case of favorable agreement this holds
usually only at a qualitative level because there is only limited knowledge
of
the actual form of the force fields governing drop-substrate systems.

Therefore, the recent report in Ref. \cite{Yuan2010} of a ultrathin
precursor film extending in time by following a power law with exponents
varying between $\approx 1/7$ and $1/4$ is surprising. This claim is based
on MD simulations of a water drop spreading on a (eventually charged) gold
substrate. However, direct inspection of the data reveals that they actually
correspond to the spreading of the drop, which consists of a quasi-spherical
cap \textit{and a foot region} (see Fig. \ref{fig1}), and not that of a film.
[For a macroscopic drop it is indeed well documented that the base radius
expands slower than the precursor film (see Secs. \ref{par_tot} and
\ref{spon}).] What the authors observe is, at most, the formation of the foot
of the drop but not, as claimed, a precursor film. This assessment stems from
the following observations:
(i) The film in Fig. 2(a) in Ref. \cite{Yuan2010} has an extent from the
edge of the drop of only three molecules, which renders a power law
fitting meaningless. (ii) In the case of the reported ``complete wetting''
a layered film, with no evidence for a precursor, is formed
(see Figs. 1(c) and 2(c) in Ref. \cite{Yuan2010}, for electric field
$E > E_s$), while nevertheless Fig. 2(c) therein features an exponent $n(E)$ stemming
from a power-law fitting of the extent of a non-existent precursor. (iii) The
droplet is too small to act as a particle reservoir for the film. The
transverse cross section of the initial droplet contains roughly $120$
molecules so that a film even with a very small lateral extent (e.g., of 15
molecules measured from the edge of the droplet) would already consume
25$\,\%$ of the initial droplet mass. A closer look at the data reveals
additional issues associated with the simulated system. After what the
authors call full spreading, the ``water''
\footnote[3]{
\label{foot12}
The quotation marks serve as a reminder that in the MD
simulations assumptions have to be made about the interaction potentials
and about the structure of the liquid, e.g., the formation and the strength
of hydrogen bonds characteristic for actual water.
} droplet shown in Fig. 1(b) in Ref.
\cite{Yuan2010} exhibits a non-zero (and rather large) contact angle, which
is confirmed by Fig. 2(c) therein listing a ``partial wetting'' state for fields
$E < E_s$. This is in conflict with the experimental evidence that a smooth
and clean gold surface is completely wetted by pure water \cite{Smith1980}.
Therefore either the interaction potentials used are not appropriate for
capturing the behaviour of the system at macroscopic scales, or the potentials
are satisfactory but the wetting behaviour of a nanoscale drop of water is
different from that of a macroscopic one in a way which is captured, rather
than bypassed, by the MD simulations. In either case, extrapolating the
results of these MD simulations towards claims of ``resolving the
Huh-Scriven paradox'' (which occurs only within the framework of
macroscopic hydrodynamics)
\footnote[4]{
\label{foot13}
The Huh-Scriven paradox is that imposing the no-slip boundary
condition at the solid-liquid interface in the case of a three-phase contact
line moving over a flat and smooth solid leads to a logarithmic divergence of
the rate of viscous dissipation in the fluid as function of, e.g., the vanishing slip length (see
Ref. \cite{deGennes1985}).
} and of understanding actual electro-wetting experiments is unwarranted.

\section{ \label{sec_new} Recent developments}

The rapid development of micro- and nano-fluidics devices and the
emergence of more accurate and faster methods of surface characterization
and analysis has recently led to a renewed interest in the spreading of
small droplets and in the formation and structure of precursor films. For
example, a new method for studying the precursor dynamics, called
epifluorescence, has been proposed. This is a microscopy technique in which
the visible light in the microscope eyepieces is the one emitted, through
fluorescence, by the specimen itself upon being exposed to light source with
a specific wavelength. It has been applied for a silicone oil (PDMS) drop
spreading on a glass substrate \cite{Hoang2011}. The advantages of this
method are a lateral resolution of ca. $0.2 ~\mu\mathrm{m}$, which is
significantly better than that of ellipsometry, and a fast response allowing
for a temporal resolution of ca. $200$ ms. In principle such a temporal
resolution  allows one to study the microscopic details of spreading even
of low
viscosity liquids, which so far has not been possible. A potential drawback
of this method is the need of enriching the liquid drop with a fluorescent
dye. This may restrict the choice of liquids, and it may even affect the
spreading, if for any reason the dye behaves as a surfactant at the
liquid-solid or the liquid-gas interfaces of interest. However, these are technical
challenges rather than basic drawbacks and it seems reasonable to expect
that an adequate choice of dyes can be found for a variety of liquids.

Tapping mode atomic force microscopy (AFM) has also been employed in recent
studies of precursor films \cite{Xu2004}. This technique, which provides a
lateral resolution down to a few nm while preserving a vertical spatial
resolution of less than one nm, had been used previously to study the
spreading of drops \cite{Villette1997,Glick1997}. By employing
a well suited
setup, with the AFM repeatedly scanning a region at a certain distance from
the macroscopic edge of a quasi-stationary drop, Xu \textit{et al} have been
able to detect and monitor a precursor film of a complex polymer
[poly(tert-butyl acrylate) brush molecules] spreading on a planar highly
oriented pyrolytic graphite substrate (which is graphite with an angular
spread between
the graphite sheets of less than 1$^\circ$).
The observed dynamic behaviour obeys the generic $\sqrt{t}$ dependence
\cite{Xu2004}. In contrast, the report based on tapping mode AFM of what is
called in Ref. \cite{Glynos2011} a ``precursor'' film of nanometer extent at
the periphery of a small drop consisting of star-shaped
polymers, in contact
with a silicon oxide substrate has to be treated with caution. The set-up of
this experiment is that of a \textit{dewetting} situation and, as the
authors acknowledge, the actual film observed was the one remaining on the
surface during dewetting, rather than one emerging from the drops. Therefore,
it is likely an equilibrium film instead of a precursor.

The ever increasing computing power and storage capabilities of computers combined 
with the development of new, efficient methods and
algorithms for numerical simulations allow one to explore new types of computer
experiments as well as to significantly improve the accuracy of previous results.
For example, Ref. \cite{Chibbaro2008} employed both MD and Lattice-Boltzmann
(LB) simulations in order to study the capillary imbibition and precursor film 
formation in nanotubes. It has been reported that both types of simulations agree 
in that a molecularly thin precursor film forms ahead of the advancing meniscus, 
and the lateral extension of this film grows in time as $\sqrt{t}$, similarly to
the observations for other geometries. This is particularly interesting as in
this set-up the position of the meniscus moves with a similar time dependence
(i.e., following Washburn's law), but with a smaller prefactor. Therefore
this is a more complex situation, in which the dynamics of the precursor film
is more connected to that of the meniscus rather than being independent of
it. Moreover, excellent quantitative agreement has been reported between the
results of MD and LB simulations. This opens the perspective of using LB
simulations, which allow the exploration of time and length scales unaccessible 
by other numerical methods, for a variety of other systems.

Finally, we note a recently reported \cite{Douezan2011,Douezan2012} 
surprising example of the formation of a ``ultrathin'' (i.e., monolayer) 
precursor film, with a spreading dynamics described well by a $\sqrt{t}$ 
behaviour. The system is a spheroidal aggregate of cells, the surface of each 
cell containing a certain level of a cell-cell adhesion agent (E-cadherin 
molecules), in contact with a glass substrate decorated with a mixture of a 
cell-adhesion agent (fibronectin) and a non-adhesive component 
(PEG-poly-L-lysine,PEG-PLL). The behaviour of such cellular aggregates bears 
strong similarities with that of a viscoelastic liquid drop. Accordingly, a
spreading coefficient $S_0= W_{CS} - W_{CC}$, given by the difference
between the cell-cell ($W_{CC}$) and cell-substrate ($W_{CS}$) adhesion
energies per unit area, can be defined analogously to that of a liquid.
By varying the amounts of E-cadherin molecules on the cells surface and
fibronectin on the glass substrate, it was possible to change $S$ from
negative to positive values. This is the equivalent of a transition from
incomplete wetting to complete wetting for a liquid-on-solid system. This
was confirmed by the corresponding ``wetting'' behaviour of the cell
aggregates. In the complete wetting regime, in which the glass substrate was
fully covered by fibronectin, a precursor film, with a thickness equal to
one layer of cells, was observed to spread on the substrate ahead of the cell
aggregate. The structure of the film is more compact, i.e., liquid-like for
stronger cell-cell adhesion; it turns into a sparse, two-dimensional gas-like
configuration as the cell-cell adhesion is weakened. The analogy with the
liquid-on-solid systems is completed by the observation of an expansion of
the liquid-like precursor film showing good agreement with a $\sqrt{t}$
time dependence.

\section{Summary and outlook}
\label{summary}

It can be easily inferred that the combination of experiment, theoretical
analysis, and numerical simulations has led to a good understanding of the
mechanisms of the formation and the dynamics of precursor films on
homogeneous, planar, smooth substrates. However, in our view even for this
most basic case several important questions still await answers:\newline
\textbullet~ What are the necessary and sufficient conditions for a
precursor film to emerge during the spontaneous spreading of a
(quasi)non-volatile drop?  Are these the same for a solid-on-solid system
such as the metal-on-metal one discussed in
Sec. \ref{metal_metal_exper}? \newline
\textbullet~ A good understanding of the dynamics of mesoscopically
thick precursor films is available, and theoretical lattice-gas models
seem to capture well the dynamics of ultrathin films. A comprehensive
theory, which would deal with all length scales and recover the behaviour
of thick and ultrathin films as particular limiting cases remains to be
developed.\newline
\textbullet~ The early experiments, and so far the only systematic studies
of ultrathin precursors, focused on the case of complete wetting. Apart
from Refs. \cite{Tiberg2,Tiberg3}, we are not aware of a similar systematic
experimental study of the partial wetting case. Such results are needed to
elucidate the importance of the detailed shape of inter-particle pair
potentials
as revealed by the MD simulations. As we discussed in Secs. \ref{sec_MD_drop}
and  \ref{sec_MD_metal}, the choice of the interaction potentials leads to
conflicting results on whether or not a ultrathin precursor film emerges in
partial-wetting situations.\newline

By using KMC simulations of lattice gas models few studies have
addressed the issue of ultrathin precursor films spreading on chemically
inhomogeneous substrates \cite{val1,val2,Pesheva2002} or upon encountering
chemical patterns \cite{Popescu2003}. But questions such as how precursor
films form and spread on chemically inhomogeneous or patterned substrate are
yet to be explored systematically. Millimeter-sized drops which are set
in motion upon contact with a chemically patterned substrate have been recently
studied experimentally \cite{Bliznyuk2011}. However, the question of formation
and dynamics of precursor films, as well as that of their influence on the
emerging motion of the drop, was not yet addressed.
In thermal equilibrium partially wetting nanodrops coexist with a microscopic
film. When placed near topographic or chemical heterogeneities on the substrate
surface, they interact with them via this film and therefore they are set into
motion \cite{Ali2006,Ali2008_la,Ali2008_jcp,Rauscher2009_sm,Ali2009}.
This naturally leads to the question of how these dynamic phenomena translate
to the case in which a ultrathin precursor film, which spreads ahead of a
nanodrop, will interact with such geometrical or chemical structures. For
example, one can consider the encounter of a precursor with a small chemical
patch, which was addressed briefly in Ref. \cite{Popescu2005},
and how in turn this interaction has an influence on the drop. We note here
a recently reported elegant example of such precursor-film mediated
interactions, applied to achieve thermocapillary actuation even at microscopic
scales \cite{Zhao2011}. By applying a thermal gradient to a parylene-coated
silicon substrate, a ultrathin precursor film emerges out of a heptanol drop.
This film mediates both the driving of this oil drop towards a water drop
sitting nearby, as well as the encapsulation of the water drop by the heptanol.
Once the encapsulation is complete, the composite starts moving along the
temperature gradient, and thermocapillary actuation is achieved.

The complexity increases by considering whether or not precursor films
occur and, if so, how they spread in situations in which the substrate
is responsive. Examples are soft substrates such as in the case of a liquid
lens at the interface
between two immiscible liquids, or at flexible, hairlike substrate structures.
Spreading on curved substrates, such as for a sessile drop on a cylindrical
or conical rod \cite{Lv2012}, as well as on substrates with other types of
complex geometries, are issues of significant interest which have barely been
studied. For example, the very interesting question of
spreading of a drop and the formation of precursor films inside a wedge (corner)
have been recently approached using MD simulations \cite{Zhao2012}. These
preliminary results (although suffering from a premature interpretation
in terms of a heuristic model presumably applicable to the motion of a three-phase
contact line but not to the dynamics of ultrathin precursor films \cite{Blake})
hint at a wealth of qualitatively new phenomena, such as the
coexistence of two-dimensional and one-dimensional precursor films spreading ahead
of the ``macroscopic'' contact line.
The occurrence and dynamics of precursor films in situations where a volatile
or non-volatile colloidal suspension spreads on a substrate \cite{Maki2011}
provide further important challenges for theory and for applications of
wetting.

\addcontentsline{toc}{section}{Acknowledgements}
\section*{Acknowledgements}

MNP and SD acknowledge financial support from the Australian Research
Council (grant DP1094337). AMC wishes to thank Prof. John
Ralston for many helpful discussions.

\addcontentsline{toc}{section}{References}
\section*{References}

\bibliographystyle{unsrt}
\bibliography{refs_precursor_film_v10} 


\end{document}